%% file: AnonymousSubmission2027.tex
\documentclass[letterpaper]{article}
\usepackage{aaai2027}  

\nocopyright  

\usepackage[hyphens]{url}
\usepackage{graphicx}
\usepackage{natbib}
\usepackage{caption}
\usepackage{amsmath}
\usepackage{amssymb}
\usepackage{amsfonts}
\usepackage{algorithm}
\usepackage{algorithmic}

\usepackage{newfloat}
\usepackage{listings}
\DeclareCaptionStyle{ruled}{labelfont=normalfont,labelsep=colon,strut=off}
\floatstyle{ruled}
\newfloat{listing}{tb}{lst}{}
\floatname{listing}{Listing}

\usepackage{booktabs}
\usepackage{multirow}

\title{GCNO: Gramian Chebyshev Neural Operator for Physics-Based Compression of Wireless Channels}

\author{
    Rafid Umayer Murshed,\textsuperscript{\rm 1}
    Shahab Hamidi-Rad,\textsuperscript{\rm 2}
    Elahe Soltanaghai,\textsuperscript{\rm 1}
    Akshay Malhotra\textsuperscript{\rm 2}
}
\affiliations{
    \textsuperscript{\rm 1}Department of Computer Science, University of Illinois Urbana-Champaign\\
    \textsuperscript{\rm 2}InterDigital AI Lab\\
    \texttt{\{rum3, elahe\}@illinois.edu}, \texttt{\{Shahab.Hamidi-Rad, akshay.malhotra\}@interdigital.com}
}

\begin{document}

\maketitle

\begin{abstract}
Large antenna arrays allow wireless systems to serve more users and achieve higher data-rates, but they also make channel feedback expensive: the receiving device must repeatedly report a large complex-valued channel matrix to the base station. Most neural compressors treat this matrix like an image and replace it with a fixed-length code that only a matched neural decoder can interpret. The message therefore does not adapt to channel complexity, and changing the antenna count typically requires retraining. We ask whether a device can instead report only the few dominant propagation paths underlying each channel. We introduce the \textbf{Gramian Chebyshev Neural Operator (GCNO)}, a physics-based, variable-rate compressor that identifies a sample-dependent set of path directions. GCNO uses receive–transmit channel structure to locate paths, a first-order Taylor correction to refine directions that fall between grid points, and least squares to recover their complex strengths. It is trained without path labels, and the base station reconstructs the channel analytically from the transmitted path tuples rather than through a learned decoder. Across three ray-traced environments, GCNO achieves better reconstruction accuracy at the same payload—or lower payload at the same accuracy—than neural feedback baselines, and transfers to unseen antenna counts without retraining.
\end{abstract}


\input{Sections/Intro}

\input{Sections/Related_Works}

\input{Sections/Background}

\input{Sections/Methodology}

\input{Sections/results}

\section{Conclusion}
\label{sec:conclusion}
This paper examined whether a wireless channel is better compressed through its dominant propagation paths than through a fixed-size code produced for a paired neural decoder. Across ASU, Dallas, and Seattle, the results support the path-based view. GCNO finds a channel-dependent set of directions, Taylor refinement moves them away from the fixed grid, and least squares estimates their strengths. The device reports the retained path tuples. This variable payload gives a better trade-off between reconstruction accuracy and feedback size than neural compressors, while the physical reconstruction rule allows the same model to work when the antenna count changes. The method is trained from channel reconstruction alone, without path labels.

Our experiments focus on narrowband spatial channels. We sought to isolate the value of path-based compression and compare it fairly with encoder-decoder methods, rather than combine that question with delay and OFDM modeling. Extending the representation to wideband channels requires adding delay, or an equivalent frequency coordinate, to each path. Path directions and delays remain nearly constant across nearby subcarriers, and gain changes follow a structured pattern. One path description can therefore serve many frequencies. This shared structure may make OFDM channels even more compressible than the narrowband channels studied here.

\bibliography{aaai2027}

\input{Sections/Appendix_A}
\input{Sections/Appendix_B}
\input{Sections/Appendix_C}


\end{document}

%% file: Sections/Intro.tex
\section{Introduction}

Modern cellular systems place large antenna arrays at the base station, known as massive MIMO technology. These arrays can serve many users simultaneously and carry far more data, but only when the base station knows the current wireless channel: how the transmitted signal fades, reflects, and combines on its way to the user. How the base station obtains this information depends on the duplexing scheme. In frequency-division duplex systems, the uplink and downlink occupy different frequency bands, so the user measures the downlink channel state information (CSI) and reports it to the base station over a control link. Reporting CSI in full is costly. The channel between an $N_r \times N_t$ array pair is a complex matrix that requires $2N_rN_t$ real numbers to describe. A $32 \times 32$ channel already requires $2{,}048$ numbers, and this cost continues to grow as antenna counts increase toward 6G~\cite{marzetta2010,larsson2014}. Because this feedback consumes control-link resources that could otherwise carry data, CSI compression has become a central problem in massive MIMO~\cite{larsson2014}. This raises the main question of our work: \textit{What compact channel representation can reduce CSI feedback without sacrificing reconstruction accuracy or tying the method to one antenna configuration?}

Deep learning (DL) has been studied extensively for compressing CSI feedback~\cite{oshea2017,guo2022overview}. Most learned approaches treat the channel matrix similarly to an image. A neural encoder on the user device (UE) compresses the matrix into a short code, which a matched neural decoder at the base station (BS) uses to reconstruct the channel~\cite{csinet}. Subsequent methods improve this architecture through convolutional, recurrent, and attention-based processing~\cite{transnet,xu2021transformer,dcrnet,csinet_lstm}, while retaining the same paired encoder--decoder structure. This structure has two important limitations. Channels with different levels of complexity are compressed into codes of the same length, preventing the feedback overhead from adapting to the amount of information in each channel. In addition, the encoder and decoder are jointly trained for a specific antenna configuration, so changing the antenna count generally requires both networks to be retrained. These limitations motivate two corresponding evaluation criteria. We assess how few values a method must transmit to achieve a target channel-reconstruction normalized mean-squared error (NMSE) and how accurately it reconstructs channels with different antenna counts without retraining.

To address the limitations of prior work, we represent the channel through its underlying multipath structure rather than an antenna-specific latent code. Many outdoor and high-frequency channels are dominated by only a few strong propagation paths~\cite{akdeniz2014millimeter,elayach2014spatially}. Under the geometric channel model, each path is described by its complex gain and transmit and receive directions, and the full channel can be reconstructed from the corresponding array responses~\cite{alkhateeb2014}. A channel with $K$ paths therefore requires only $4K$ real-valued parameters, compared with $2N_rN_t$ values for the full complex matrix. The user feeds back these path parameters, and the base station reconstructs the channel analytically without a matched neural decoder. Because the underlying paths are determined by the propagation environment rather than the antenna count, the same representation can be used across different antenna configurations, while its feedback length naturally follows the number of dominant paths. 

\begin{figure}[t]
    \centering
    \includegraphics[trim= 1cm 0.5cm 0.1cm 1.1cm, clip, width=1.0\columnwidth]{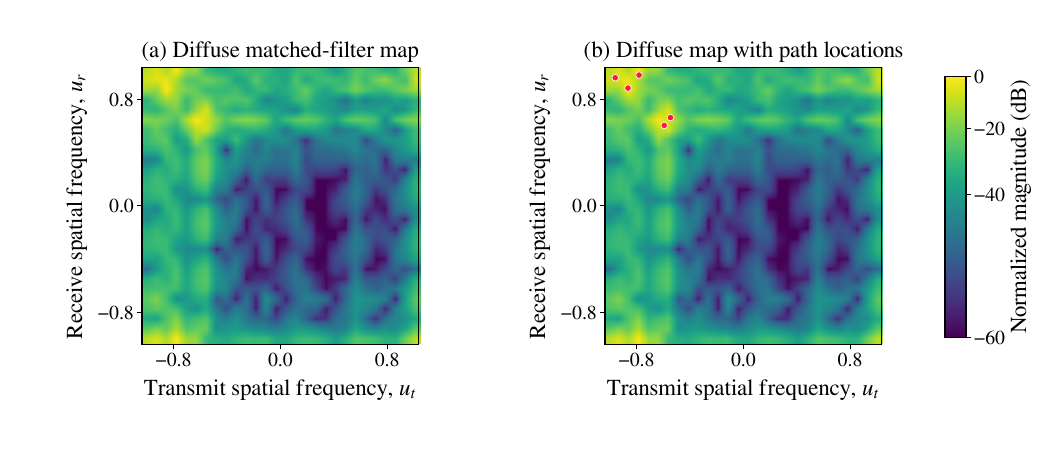}
    \caption{Off-grid paths spread their energy across neighboring cells. The map is shown (a) alone and (b) with the underlying paths marked, illustrating why a grid-only description may use several entries for one physical path.}
    \label{fig:offgrid}
\end{figure}

Having reduced the feedback problem to a few path parameters, the main challenge is to estimate the continuous transmit and receive directions of those paths. Classical estimators search over a fixed grid of candidate transmit and receive directions and identify the grid points that best match the measurements~\cite{klukas1998line}. Such a grid is unsuitable as the final feedback representation because a physical path rarely aligns exactly with one grid point. Its energy instead spreads across neighboring bins (fig.~\ref{fig:offgrid}), and several grid coefficients may be needed to represent a single path, especially when patterns from multiple paths overlap~\cite{tang2013offgrid}. This would increase the feedback length and undermine the compact path-based representation. Directly predicting continuous directions with a neural network is also ill-posed because small directional changes produce rapidly oscillating phase variations in the array response, resulting in a highly non-convex learning objective~\cite{wagle2025physics}. We therefore adopt a hybrid approach that uses the direction grid only to construct a structured evidence map and learns to map it to one continuous direction pair for each path.

Recovering continuous path directions across different antenna configurations requires a model whose learned parameters are not tied to a fixed input dimension. This motivates the use of neural operators, which learn transformations that can be evaluated on inputs of different sizes~\cite{fno,deeponet}. However, the grid-matching pattern also varies across channels according to the locations and interactions of their paths. Standard neural operators apply filters defined by a fixed basis or graph and therefore process every channel using the same filtering structure~\cite{fno,chebnet}. To address this challenge, we introduce the \emph{Gramian Chebyshev Neural Operator} (GCNO) to retain the size flexibility of neural operators while adapting the filtering to each observed channel. GCNO constructs receive- and transmit-side Gramian matrices from the channel and uses them to define its Chebyshev filters. The resulting operators reflect the current channel structure, while the learned polynomial coefficients remain shared across antenna configurations.

Our complete approach follows a simple division of work for recovering the path angles and corresponding gains. GCNO first identifies likely path directions (angles) from the channel. Continuous refinement then moves each selected direction away from its grid point and toward a more accurate location. GCNO retains only paths that provide meaningful reconstruction improvement, allowing the path count $K$ to adapt to the channel. Once the path directions are extracted, a least-squares (LS) solver computes their complex strengths (gains). The model learns from channel reconstruction alone and does not require ground-truth path directions, strengths, or path counts. At inference time, the user device feeds back only the recovered path tuples, and the base station reconstructs the full channel by combining the reported path gains and angles with the transmit and receive array geometries.

We evaluate GCNO on city-scale ray-traced channels from Arizona State University (ASU), Dallas, and Seattle. Across the full NMSE--payload trade-off, GCNO achieves lower NMSE at the same payload, or lower payload at the same NMSE, when compared with eight neural feedback baselines. It also retains substantially more accuracy when evaluated on antenna counts not seen during training. Replacing GCNO with CNN, Fourier, polynomial, or dilated-convolutional alternatives weakens this transfer or the NMSE--payload trade-off, showing that the gain does not come only from model size. Finally, retraining the model without continuous refinement clearly worsens NMSE, confirming that correcting off-grid directions is necessary for compact path feedback. Our contributions are:

\begin{itemize}
    \item \textbf{Variable-rate path compression.} We formulate CSI feedback as a variable-length list of retained paths, requiring $4K$ real values for $K$ reported paths. The feedback size therefore adapts to the channel, and the known array model reconstructs the channel without a matched neural decoder. Across three environments, this framework provides better reconstruction at the same feedback size, or uses fewer reported values at the same accuracy.

    \item \textbf{Gramian Chebyshev Neural Operator.} We introduce GCNO, which constructs channel-specific filters from receive- and transmit-side correlations while sharing its learned coefficients across antenna counts. This design improves both path recovery and generalizes to antenna counts not seen during training.

    \item \textbf{Continuous refinement and LS recovery.} We refine each selected receive--transmit direction pair within its grid cell and recover its gain through LS, without using
path labels. A separately retrained grid-only variant, evaluated under
the same support-selection and payload rules, performs substantially
worse in reconstruction and angular accuracy across all three
environments.

\end{itemize}

%% file: Sections/Related_Works.tex
\section{Related Work}

Our work draws on learned CSI feedback, neural operators, and sparse off-grid recovery.

\noindent\textbf{Neural CSI feedback and learned compression.}
CsiNet~\cite{csinet} introduced the common encoder--decoder design: the user compresses the real and imaginary parts of CSI into a dense code, and a paired base-station network reconstructs the channel. Later work improves this design through residual or dilated convolutions~\cite{crnet,dcrnet}, stronger CsiNet variants~\cite{csinet_plus}, recurrent refinement~\cite{csinet_lstm}, attention~\cite{attention_csi}, and transformers~\cite{xu2021transformer,transnet}. These methods still use a fixed-length code tied to a trained decoder, so feedback does not naturally follow channel complexity, and changing the antenna configuration usually requires a matched model. We instead send a variable set of path gains and effective spatial directions that the receiver converts to CSI using the array geometry.

\noindent\textbf{Neural operators and spectral filters.}
Neural operators, including FNO~\cite{fno} and DeepONet~\cite{deeponet}, learn maps between sampled functions and can operate across discretizations. FNO uses a fixed periodic Fourier basis, while Chebyshev graph networks~\cite{chebnet} filter a predefined graph Laplacian. These filtering structures do not adapt to an individual channel, and Fourier periodicity is poorly matched to a bounded angular field of view. GCNO instead forms receive- and transmit-side Gramians from each sample and applies shared Chebyshev polynomial filters, enabling channel-dependent processing across array sizes.

\noindent\textbf{Sparse recovery and off-grid methods.}
OMP~\cite{omp}, basis pursuit~\cite{basis_pursuit}, and sparse Bayesian learning~\cite{sbl} recover dictionary supports without path labels, but off-grid paths spread across several atoms. Atomic-norm recovery~\cite{tang2013offgrid}, ESPRIT~\cite{esprit}, off-grid Bayesian learning~\cite{offgrid_sbl}, and Newton refinement~\cite{newton_refine} address continuous directions but require spectral or iterative solves. LISTA~\cite{lista} and related unrolled methods~\cite{unrolled_sparse} reduce this cost, yet often remain grid-dependent or supervised. GCNO predicts support and Taylor offsets in one pass, followed by a small least-squares gain solve.

%% file: Sections/Background.tex
\section{Problem Formulation}
\label{sec:problem}

\begin{figure*}[t]
    \centering
    \includegraphics[trim=0.0cm 0.0cm 0.0cm 0.0cm, clip, height = 4.2 cm, width=\textwidth]{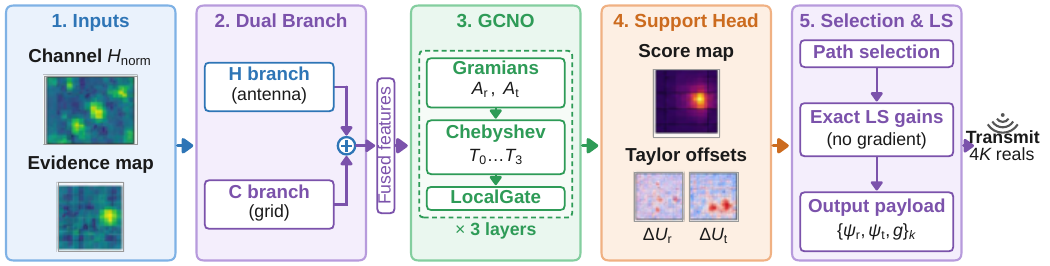}
    \caption{Overview of GCNO compression. The channel and evidence map are encoded jointly; GCNO predicts path scores and off-grid corrections, path selection retains useful directions, and LS recovers their gains. Appendix B provides more detail.}
    \label{fig:GCNO_archi}
\end{figure*}

Given a channel matrix, our goal is to describe it with as few
reported values as possible while preserving accurate reconstruction at the
BS. We do this by extracting a small set of dominant propagation
paths that captures the important structure of the channel. Each retained path
is described by one complex gain and two spatial directions. We use \(K\) to
denote the number of paths retained for a particular channel. The compression
problem is therefore to choose both \(K\) and the corresponding path parameters
so that the feedback remains small and the reconstructed channel remains
accurate. We also analyze whether the same learned compressor can be applied when
the antenna count changes without retraining.

\subsection{Physical Path Representation}

A propagation path produces a predictable phase pattern across each antenna
array. We call this pattern the \emph{array response}. The geometric channel
model represents the full channel as the sum of the contributions produced by
its paths. For the uniform linear arrays considered here, a direction is described by its
spatial coordinate \(u\in[-1,1]\). The corresponding normalized array response
is
\begin{equation}
\footnotesize
\begin{aligned}
    \mathbf{a}_{N}(u)[n]
    &=
    \frac{1}{\sqrt{N}}
    \exp\left(
        \mathrm{j}\pi
        \left(n-\frac{N-1}{2}\right)u
    \right), \\
    &\qquad n=0,\ldots,N-1.
\end{aligned}
\label{eq:steering-vector}
\end{equation}
We report \(u\) through the effective spatial angle
\(\psi=\arcsin(u)\). This is the direction seen by the array.

Let \(K^\star\) denote the total number of paths in the standard geometric
channel model~\cite{alkhateeb2014}. The measured channel is written as
\begin{equation}
    \mathbf{H}
    =
    \sum_{k=1}^{K^\star}
    g_k\,
    \mathbf{a}_{N_r}(u_{r,k})
    \mathbf{a}_{N_t}(u_{t,k})^{\mathsf{H}},
    \label{eq:channel-model}
\end{equation}
where
\(\mathbf{H}\in\mathbb{C}^{N_r\times N_t}\),
\(g_k\in\mathbb{C}\) is the complex gain of path \(k\), and
\(u_{r,k}\) and \(u_{t,k}\) are its receive and transmit spatial directions.
Although a channel may contain many weak contributions, most of its energy is
often concentrated in a much smaller set of dominant paths. In outdoor and
high-frequency channels, this dominant set can often contain five or fewer
paths~\cite{akdeniz2014millimeter,elayach2014spatially}. Our compressed
representation therefore retains \(K\leq K^\star\) paths that preserve the
important channel energy. Here, \(K\) is the number of paths used for
compression, not the total number of physical paths in the environment.

\subsection{Path-Based Feedback and Reconstruction}

Our approach frames CSI feedback directly as a list of retained path
parameters. For each path, the device reports its complex gain and its receive
and transmit effective spatial angles. The transmitted description is
\begin{equation}
    \mathcal{T}(\mathbf{H})
    =
    \left\{
    \left(
        \Re\widehat g_k,\,
        \Im\widehat g_k,\,
        \widehat\psi_{r,k},\,
        \widehat\psi_{t,k}
    \right)
    \right\}_{k=1}^{K}.
    \label{eq:path-list}
\end{equation}
Each path contributes four real values, so the feedback size is \(4K\).
Because \(K\) is selected separately for each channel, a channel that can be
represented accurately with fewer paths requires fewer reported values. Appendix~A theoretically derives when $4K$ feedback is more compact than a fixed latent.

At the base station, the reported effective angles are converted back to spatial coordinates: $\widehat u_{r,k}=\sin(\widehat\psi_{r,k})$ and $\widehat u_{t,k}=\sin(\widehat\psi_{t,k})$. The channel is then reconstructed through the same channel model as
\(\widehat{\mathbf H}=\sum_{k=1}^{K}\widehat g_k\,
\mathbf a_{N_r}(\widehat u_{r,k})
\mathbf a_{N_t}(\widehat u_{t,k})^{\mathsf H}\).
Each retained path still requires only four reported values when the antenna
count changes. The base station simply evaluates the same array-response
formula for the new array size. Thus, the feedback format and reconstruction
rule do not depend on a neural decoder trained for one fixed antenna
configuration.

%% file: Sections/Methodology.tex
\section{GCNO for Physics-Based CSI Compression}
\label{sec:method}

The problem formulation represents the channel using a compact set of path directions and complex gains. We now describe how these parameters are recovered from the measured channel. The pipeline first compares the measured channel with a dictionary of candidate path responses constructed from the known antenna geometry. This produces a direction-space map indicating which receive and transmit directions are most consistent with the observation. GCNO takes the measured channel and this map as inputs and outputs a score for each candidate direction together with continuous corrections that move the estimated directions beyond the discrete grid. The resulting candidates are then filtered to retain distinct paths that contribute meaningfully to channel reconstruction, allowing the estimated path count to vary across channels. Given the retained directions and the measured channel, LS finds the corresponding complex gains.

During training, path selection is replaced by a differentiable approximation so that the complete pipeline can be optimized using channel reconstruction error. The corrected directions are converted into candidate channel responses, least squares estimates their gains, and the retained responses are combined to reconstruct the measured channel. The reconstruction loss therefore trains GCNO without requiring ground-truth path directions, gains, or path counts.

\subsection{Direction Grid, Dictionary, and Evidence}
\label{sec:physical_evidence}

The entries of \(\mathbf H\) contain the complete channel observation, but they
do not directly show which propagation directions produced it. We obtain a more
useful view by first testing a finite set of possible directions. Let
\(\{u^r_i\}_{i=1}^{R_r}\) and \(\{u^t_j\}_{j=1}^{R_t}\) be evenly spaced receive
and transmit directions over the field of view. Combining these two lists gives
a two-dimensional \emph{direction grid}. Each cell \((i,j)\) represents one
candidate receive--transmit direction pair.

For every cell, we use the known array responses from
Equation~\eqref{eq:steering-vector} to precompute the channel pattern created by
a unit-gain path at that pair:
\begin{equation}
\footnotesize
    \mathbf D_{ij}
    =
    \mathbf a_{N_r}(u^r_i)
    \mathbf a_{N_t}(u^t_j)^{\mathsf H},
    \qquad
    \mathbf H
    \approx
    \sum_{i=1}^{R_r}\sum_{j=1}^{R_t}
    W_{ij}\mathbf D_{ij}.
    \label{eq:grid-dictionary}
\end{equation}
We call \(\mathbf D_{ij}\) a \emph{dictionary atom}, and the fixed collection
of all such atoms the \emph{dictionary}. The second relation is exact when all
paths fall on the grid and becomes an approximation otherwise. In that
relation, \(W_{ij}\) is the complex contribution assigned to cell \((i,j)\).

This construction follows the dictionary-based linearization of~\cite{wagle2025physics,wagle2025physics_workshop}. Predicting
continuous directions directly places oscillatory array-response functions
inside backpropagation and creates a difficult non-convex objective. Holding
the candidate patterns fixed instead makes the reconstructed channel a linear
combination of known matrices. In our method, the grid is only an intermediate
search space; its cells are not the final reported paths.
To reveal which cells agree with the measured channel, we compute the complex
inner product \(C_{ij}=\langle \mathbf D_{ij},\mathbf H\rangle_F\) between
\(\mathbf H\) and each dictionary atom. The resulting matrix \(\mathbf C\) is
the \emph{evidence map}. A large
\(\lvert C_{ij}\rvert\) means that the channel contains a component consistent
with that direction pair. An off-grid path usually produces a broad pattern
over several nearby cells rather than one isolated value. We therefore give
GCNO both the complete observation \(\mathbf H\) and the evidence
\(\mathbf C\). Separate lightweight encoders process their real and imaginary
parts and combine them into a common complex feature field.

\subsection{Finding Likely Paths with GCNO}
\label{sec:gcno}

The evidence map indicates where paths may lie, but several paths can produce
overlapping patterns, and one off-grid path can activate many cells. GCNO is
designed to interpret this full receive--transmit structure rather than process
each cell independently. Its central idea is to measure how the current
features are related along the receive and transmit axes, then use those
sample-specific relationships to filter the features.

Let
\(\{\mathbf X^{(\ell)}_c\}_{c=1}^{C_\ell}\) denote the complex feature maps at
layer \(\ell\). GCNO forms two correlation matrices, called Gramians:
\begin{equation}
\footnotesize
    \mathbf G^{(\ell)}_r
    =
    \sum_{c=1}^{C_\ell}
    \mathbf X^{(\ell)}_c
    \mathbf X^{(\ell)\mathsf H}_c,
    \qquad
    \mathbf G^{(\ell)}_t
    =
    \sum_{c=1}^{C_\ell}
    \mathbf X^{(\ell)\mathsf H}_c
    \mathbf X^{(\ell)}_c .
    \label{eq:gcno-gramians}
\end{equation}
The first records relationships across receive positions, while the second
records relationships across transmit positions. After normalization, they
give operators \(\mathbf A^{(\ell)}_r\) and
\(\mathbf A^{(\ell)}_t\). GCNO then filters each feature map using low-order
Chebyshev polynomials (We use order upto Q = 3):
\begin{equation} 
\footnotesize
    \mathbf Y^{(\ell)}_c
    =
    \sum_{c'=1}^{C_\ell}
    \sum_{p,q=0}^{Q}
    \Theta^{(\ell)}_{cc'pq}\,
    T_p\!\left(\mathbf A^{(\ell)}_r\right)
    \mathbf X^{(\ell)}_{c'}
    T_q\!\left(\mathbf A^{(\ell)}_t\right).
    \label{eq:gcno-filter}
\end{equation}
The matrices \(T_p(\mathbf A_r)\) and \(T_q(\mathbf A_t)\) mix information
through increasingly broad receive- and transmit-side relationships. The
coefficients \(\Theta^{(\ell)}_{cc'pq}\) are learned, but the Gramians are
recomputed from every input channel. Thus, GCNO uses the same learned rule for
all samples while adapting how that rule is applied to each channel. We use
three layers and the polynomials \(T_0,\ldots,T_3\); normalization, recurrence,
and local residual corrections are detailed in Appendix~B.

The output head produces three maps. The score \(S_{ij}\) gives the priority
for testing a path near grid cell \((i,j)\), while
\(\Delta U^r_{ij}\) and \(\Delta U^t_{ij}\) correct its receive and transmit
directions. GCNO does not predict path gains. Its learned coefficients depend
on feature channels and polynomial orders rather than particular antenna
indices. When the antenna count changes, the Gramians and physical patterns
change size, but the same learned filtering coefficients can still be used.

\subsection{Off-Grid Refinement, Path Selection, and LS}
\label{sec:path_selection}

A grid cell gives only a coarse direction estimate. GCNO refines it within the
cell as
\begin{equation}
\begin{aligned}
    \widehat u^r_{ij}
    &=
    u^r_i+\Delta U^r_{ij},
    &
    \widehat u^t_{ij}
    &=
    u^t_j+\Delta U^t_{ij},\\
    \mathbf B_{ij}
    &=
    \mathbf D_{ij}
    +\Delta U^r_{ij}\mathbf D^r_{ij}
    +\Delta U^t_{ij}\mathbf D^t_{ij}.
\end{aligned}
\label{eq:taylor-refinement}
\end{equation}
Each correction is bounded to half a grid interval on either side of the atoms. The fixed derivative atoms
\(\mathbf D^r_{ij}\) and \(\mathbf D^t_{ij}\) describe how the path pattern
changes locally with receive and transmit direction. Thus,
\(\mathbf B_{ij}\) is a first-order Taylor approximation of the pattern at the
corrected direction. The dictionary and both derivative dictionaries are
analytical, non-trainable tensors. Training therefore does not differentiate
through newly generated steering functions. Appendix~B gives their
construction and the approximation error.

The score map may contain more candidates than should be reported. At
inference, our approach examines cells in descending score order. A candidate is
ignored if its corrected pattern is nearly identical to a path already retained.
Otherwise, it is temporarily added and all complex gains are refitted jointly
by ridge LS. The candidate is retained only when this addition reduces the
channel reconstruction error by a sufficient amount. The scan ends when
further candidates no longer provide a meaningful improvement. This procedure
lets a simple channel retain fewer paths and a richer channel retain more; the
resulting number is \(K\). Exact duplicate, admission, and stopping thresholds are
given in Appendix~B.

For the final retained directions, the device forms the exact analytical path
patterns under no gradient and uses LS to find the gains that jointly best
reconstruct \(\mathbf H\). Joint fitting accounts for overlap among the retained
paths and avoids asking the neural network to estimate their complex gains.
The resulting gains and effective spatial angles form the \(4K\)-value message
in Equation~\eqref{eq:path-list}. The base station only evaluates that same geometric
channel model with known array response; it does not run GCNO
or repeat the selection procedure.

\begin{figure*}[t]
    \centering
    \includegraphics[height = 4.5cm, width=\textwidth]{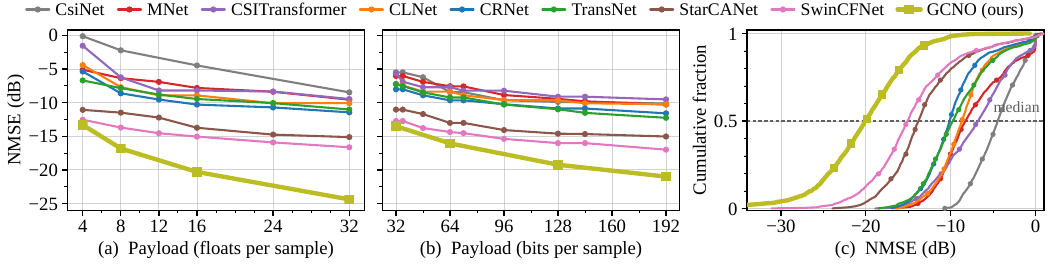}
    \caption{Rate--distortion results on ASU. Median NMSE is shown against (a) unquantized payload and (b) total feedback bits; (c) compares the test-channel error distributions at 16 transmitted values. Lower and farther left is better.}
    \label{fig:RD_Curve}
\end{figure*}

\subsection{Learning from Channel Reconstruction}
\label{sec:unsupervised_training}

Sorting candidates and making hard retention decisions are not differentiable.
During training, we replace them with a smooth version of the same process. A
fixed maximum number of soft candidates is available. Each candidate forms a
weighted average of the corrected grid locations, and nearby scores are reduced
before forming the next candidate so that different candidates cover different
regions. An activity value \(\alpha_k\in[0,1]\) controls how strongly candidate
\(k\) contributes. Therefore, \(\sum_k\alpha_k\) is a differentiable estimate
of the retained path count rather than a fixed \(K\).

The corresponding first-order patterns are built using
Equation~\eqref{eq:taylor-refinement}, and differentiable ridge LS determines
their gains jointly. Let \(\widehat{\mathbf H}_{\mathrm{soft}}\) denote this
training-time reconstruction. With \(\mathbb E_{\mathbf H}\) denoting the
average over training channels, the objective is
\begin{equation}
\footnotesize
\begin{aligned}
    \mathcal L
    ={}&
    \mathbb E_{\mathbf H}
    \left[
        \log\!\left(
            \operatorname{NMSE}
            (\mathbf H,\widehat{\mathbf H}_{\mathrm{soft}})
            +\varepsilon
        \right)
        +
        \lambda_{\mathrm{rate}}\sum_k\alpha_k
    \right] \\
    &+
    \lambda_{\mathrm{dup}}\mathcal L_{\mathrm{dup}}
    +
    \lambda_{\mathrm{off}}\mathcal L_{\mathrm{off}}
    +
    \lambda_{\mathrm{score}}\mathcal L_{\mathrm{score}} .
    \label{eq:method-training-loss}
\end{aligned}
\end{equation}
The first term rewards accurate tuple-based reconstruction, while the second
discourages unnecessary paths. The remaining terms discourage repeated
candidates, corrections near the edge of a grid cell, and broadly activated
score maps. Their exact definitions and the smooth selection procedure are
provided in Appendix~B.

Every quantity used for learning comes from the observed channel, the fixed
dictionaries, the model outputs, or the LS reconstruction. Ground-truth path
directions, gains, and path counts are not used in the loss, validation
criterion, checkpoint selection, or hyperparameter selection. At deployment,
the smooth training approximation is discarded and replaced by the adaptive
hard selection and exact no-gradient LS procedure described above. Fig. \ref{fig:GCNO_archi} summarizes GCNO.

%% file: Sections/results.tex
\section{Experimental Protocol}
\label{sec:experimental_protocol}

\subsection{Datasets and Splits}
\label{sec:datasets_splits}

We evaluate on the ASU, Dallas, and Seattle ray-traced DeepMIMO
scenarios~\citep{alkhateeb2019deepmimo}. ASU represents a campus;
Dallas and Seattle provide distinct urban geometries. Holding channel
generation fixed while changing the scene tests whether our findings
depend on one map. For each scenario, we use disjoint sets of 10,000
training, 2,000 validation, and 1,500 test channels and train a separate
model. The test set remains untouched until all choices are fixed, and
path annotations are used only for final test-set diagnostics.

\subsection{Baselines}
\label{sec:baselines}

We compare GCNO with eight paired encoder--decoder models:
CsiNet~\citep{csinet}, CRNet~\citep{crnet},
CLNet~\citep{ji2021clnet}, CSITransformer~\citep{xu2021transformer},
TransNet~\citep{transnet}, M-Net~\citep{yu2023mnet},
SwinCFNet~\citep{cheng2023swincfnet}, and
StarCANet~\citep{zhao2026starcanet}. This set spans convolutional,
multi-resolution, complex-input, MLP, global-attention,
windowed-attention, and compact designs. Each baseline sends a learned
latent code that a matched neural decoder converts back to CSI,
providing a direct comparison with our sparse physical payload and
analytic reconstruction. We retain the published architectures and
retrain them from scratch on our splits, using released implementations
when available. Appendix~C additionally reports comparisons with classical non-neural
methods and model sizes: GCNO has only $95$K trainable parameters, whereas the
closest performing neural baseline SwinCFNet exceeds 10 Millions.

\subsection{Metrics}
\label{sec:metrics}

Our primary metric is NMSE, reported in
decibels (dB). It measures reconstruction error relative to the channel energy.
For GCNO, reconstruction uses only the transmitted path tuples;
for each baseline, it uses the paired neural decoder. We report median NMSE
and its empirical CDF. Unquantized rate is the number of transmitted real
values: mean $4K$ for GCNO and the latent length for a baseline. Quantized rate
includes all transmitted bits, including the path-count header and quantized
tuple or latent entries. Unquantized (floats) and quantized (bits) budgets are separate operating points, not conversions of each other.
GCNO budgets are swept via the admission
threshold of Section 4; a stated budget is a test-set
mean, not a per-sample cap. We also report effective-angle errors for $\psi=\arcsin(u)$ in degrees. 

\subsection{Reproducibility Details}
\label{sec:reproducibility_details}

Appendix~B gives the dataset, optimization,
training, quantization, parameter-count, hardware, software, runtime, and selection details. Training and model selection use only
label-free reconstruction and rate; path counts, directions,
gains, oracle quantities, and test results never influence either. 



\section{Results}
\label{sec:results}
This section asks two practical questions: how much feedback is needed to reconstruct a channel accurately, and whether the same compressor remains useful when the antenna dimensions change. We first compare GCNO with established neural feedback methods across several environments, using both quantized and unquantized payloads, to determine which method gives the best reconstruction for a given communication cost. We then isolate the main parts of our design. The Taylor ablation tests whether continuous direction refinement is necessary to represent each off-grid path with one compact tuple, while the backbone ablation tests whether GCNO itself provides an advantage over more general neural architectures. Finally, we examine how the selected number of paths changes with channel difficulty, whether the transmitted directions correspond to meaningful propagation structure, and how well a model trained at one array size transfers to unseen antenna configurations without retraining. Appendix C provides extended results and analysis.

\begin{figure*}[t]
    \centering
    \includegraphics[height = 4.5cm, width=0.98\textwidth]{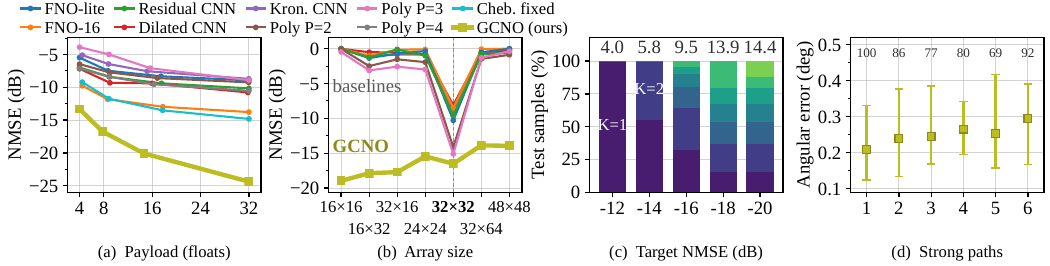}
    \caption{Ablation and Generalization results on ASU. (a) RD after replacing only the GCNO backbone. (b) Generalization of the \(32\times32\)-trained model to unseen array sizes without retraining. (c) Selected path counts (Numbers above bars show mean of $4K$) as the target NMSE becomes stricter. (d) Angular error and recall (above markers) versus the number of strong paths.}
    \label{fig:Ablation}
\end{figure*}

\subsection{Main Rate--Distortion (RD) Results}
\label{subsec:main_results}

The key question is whether a variable-length path list provides a better
accuracy--payload trade-off than a fixed neural code. Figure~\ref{fig:RD_Curve}
shows that GCNO remains on the best quantized RD frontier across
the tested bit budgets. The improvement also holds over all of the
per-channel error distribution at 16 floats
(Fig.~\ref{fig:RD_Curve}), rather than arising from a small number of easy
channels. Table~\ref{tab:crossdataset} shows the same ordering in Seattle and
Dallas at both quantized and unquantized rates. All GCNO reconstructions use
only the transmitted tuples and LS gains. These results support our
claim: physical path feedback can provide better reconstruction per
transmitted value and is not tied to one environment.

\begin{table}[t]
\centering
\caption{Median NMSE (dB; lower is better) vs 3 best baselines. Full Comparison is in Appendix C.}
\label{tab:crossdataset}
\small
\setlength{\tabcolsep}{4.5pt}
\renewcommand{\arraystretch}{1.08}
\resizebox{\columnwidth}{!}{%
\begin{tabular}{l cc cc}
\toprule
& \multicolumn{2}{c}{\textbf{Seattle}}
& \multicolumn{2}{c}{\textbf{Dallas}} \\
\cmidrule(lr){2-3}\cmidrule(lr){4-5}
Method & 64 bits & 16 floats & 64 bits & 16 floats \\
\midrule
TransNet  & $-9.71$  & $-10.11$ & $-6.01$  & $-6.22$ \\
StarCANet & $-13.04$ & $-13.65$ & $-9.35$  & $-9.79$ \\
SwinCFNet & $-14.37$ & $-14.70$ & $-10.71$ & $-11.28$ \\
\midrule
\textbf{GCNO (Ours)}
& $\mathbf{-14.93}$ & $\mathbf{-18.15}$
& $\mathbf{-15.22}$ & $\mathbf{-19.23}$ \\
\bottomrule
\end{tabular}%
}
\end{table}

\subsection{Taylor On/Off Ablation}
\label{subsec:taylor_ablation}

Path feedback is compact only when one physical path can be represented by one
tuple. Without continuous correction, an off-grid path must be approximated
from inaccurate grid directions or several neighboring entries. We test this
mechanism by retraining GCNO without Taylor offsets while keeping the support
selection and payload rules unchanged. Table~\ref{tab:taylor_ablation} shows a
large and consistent loss in all three environments, together with much less
accurate directions. Retraining therefore cannot compensate for the missing
off-grid correction. Taylor refinement is what converts coarse grid evidence
into an accurate continuous path description, preserving the payload advantage
of path-based feedback.

\begin{table}[t]
\centering
\caption{Effect of Taylor refinement, error is median (P75).}
\label{tab:taylor_ablation}
\small
\setlength{\tabcolsep}{3.5pt}
\renewcommand{\arraystretch}{1.06}
\resizebox{\columnwidth}{!}{%
\begin{tabular}{llcc}
\toprule
Dataset & Method@16 floats & Median NMSE (dB)
& Angular error ($^\circ$) \\
\midrule
\multirow{2}{*}{ASU}
  & GCNO (Taylor)     & $-20.08$ & $0.23\;(0.37)$ \\
  & GCNO (no Taylor) & $-3.84$  & $2.01\;(2.75)$ \\
\addlinespace
\multirow{2}{*}{Dallas}
  & GCNO (Taylor)     & $-19.23$ & $0.28\;(0.59)$ \\
  & GCNO (no Taylor) & $-4.14$  & $2.17\;(3.66)$ \\
\addlinespace
\multirow{2}{*}{Seattle}
  & GCNO (Taylor)     & $-18.15$ & $0.23\;(0.46)$ \\
  & GCNO (no Taylor) & $-3.29$  & $2.11\;(3.20)$ \\
\bottomrule
\end{tabular}%
}
\end{table}

\subsection{GCNO Backbone Ablation}
\label{subsec:gcno_ablation}

We ask whether the physical decoder alone explains the improvement. We
replace only the 3 native GCNO layers while retaining the same inputs, support head, Taylor
refinement, selection rule, and LS recovery. GCNO gives the strongest
RD trade-off against CNN, dilated-CNN, FNO, static
Chebyshev, and learned-polynomial replacements
(Fig.~\ref{fig:Ablation}). Since every replacement produces the same type of path
payload, the gain cannot be attributed to the analytical decoder alone. The
sample-specific receive and transmit Gramians help GCNO identify a smaller,
more useful support.

\subsection{Physical Interpretability}
\label{subsec:physical_interpretability}

Variable-rate feedback should adapt to what each channel needs rather than act
as a fixed code under another name. As the requested accuracy becomes stricter,
GCNO assigns additional paths to a growing fraction of the channels
(Fig.~\ref{fig:Ablation}). At the strictest setting, payload continues to
increase while reconstruction improves little, revealing the point of
diminishing returns. The transmitted directions also remain accurate as the
number of strong diagnostic paths increases
(Fig.~\ref{fig:Ablation}). Recall is lower for some richer channels,
showing the expected limitation of a compact representation: it preserves the
dominant resolvable structure rather than every annotated ray. Appendix C elaborates on this.

\subsection{Generalization Across Array Sizes}
\label{subsec:generalization_robustness}

Our second main question is whether the learned compressor remains useful when
the antenna count changes. Figure~\ref{fig:Ablation} applies the same
model trained at \(32\times32\) to six changed square and rectangular array
configurations, without retraining. GCNO retains strong reconstruction quality
throughout, whereas the paired neural baselines lose most of their accuracy
away from their native dimensions. This behavior is consistent with GCNO's
design: its learned polynomial coefficients are shared across array sizes,
while the physical projections and reconstruction patterns are recomputed for
the new geometry. The result supports our claim that one
trained model can transfer across antenna counts without requiring a new
matched decoder. Appendix C provides exact details on how the baselines were adapted for various arrays.

%% file: Sections/Appendix_A.tex
\appendix

\section{When Is Path-Based Feedback Minimal?}
\label{app:path_dimension}

The main paper uses $K$ for the number of paths retained by the compressor and
$K^{\star}$ for the total number of paths in the geometric model. Throughout
this appendix, $K$ retains the same meaning as in the main paper. We isolate
the ideal $K$-path component and ask how many real coordinates are required to
represent it exactly. The hats used for estimated quantities in the main
paper are omitted here because this appendix analyzes the ideal exact
representation. This is a statement about the unquantized payload used in the
main rate--distortion comparison; it is not a finite-bit entropy bound.

\subsection{Identifiable $K$-Path Channels}

Write $g_k=x_k+\mathrm{j}y_k$ and collect the path parameters as
\begin{equation}
\footnotesize
\boldsymbol{\vartheta}
=
(x_1,y_1,u_{r,1},u_{t,1},\ldots,
 x_K,y_K,u_{r,K},u_{t,K})
\in \Theta_K\subset\mathbb{R}^{4K}.
\label{eq:app_path_parameters}
\end{equation}
After imposing a fixed canonical ordering of the paths, define the synthesis
map
\begin{equation}
\Phi_K(\boldsymbol{\vartheta})
=
\sum_{k=1}^{K}
g_k\,
\mathbf a_{N_r}(u_{r,k})
\mathbf a_{N_t}(u_{t,k})^{\mathsf H},
\label{eq:app_path_synthesis}
\end{equation}
and its real vectorization
\begin{equation}
\phi_K(\boldsymbol{\vartheta})
=
\begin{bmatrix}
\Re\!\left\{\operatorname{vec}
\left(\Phi_K(\boldsymbol{\vartheta})\right)\right\}\\
\Im\!\left\{\operatorname{vec}
\left(\Phi_K(\boldsymbol{\vartheta})\right)\right\}
\end{bmatrix}
\in\mathbb{R}^{2N_rN_t}.
\label{eq:app_real_synthesis}
\end{equation}
The finite permutation ambiguity does not change the continuous dimension;
the canonical ordering only selects one representative.

\noindent\textbf{Assumption A.1 (local identifiability).}
At the channel under consideration, all retained gains are nonzero, the path
direction pairs are distinct and lie in the interior of the unaliased field of
view, and $\phi_K$ is locally one-to-one with
\begin{equation}
\operatorname{rank}
J_{\phi_K}(\boldsymbol{\vartheta})
=
4K.
\label{eq:app_identifiability}
\end{equation}
This assumption excludes coincident or locally unresolvable paths. Its four
local directions for path $k$ are generated by
$\mathbf a_r\mathbf a_t^{\mathsf H}$,
$\mathrm{j}\mathbf a_r\mathbf a_t^{\mathsf H}$,
$g_k\mathbf a_r'\mathbf a_t^{\mathsf H}$, and
$g_k\mathbf a_r(\mathbf a_t')^{\mathsf H}$, where the steering vectors and
their derivatives are evaluated at path $k$. Full Jacobian rank states that
the corresponding $4K$ real perturbations are locally independent.

\subsection{Minimum Dimension of an Exact Fixed Latent}

\noindent\textbf{Theorem A.1 (minimal exact latent dimension).}
Let
$e:\mathbb{R}^{2N_rN_t}\rightarrow\mathbb{R}^{m}$ and
$d:\mathbb{R}^{m}\rightarrow\mathbb{R}^{2N_rN_t}$ be an encoder and decoder
that are differentiable at the considered channel. Suppose that they exactly
reconstruct every identifiable $K$-path channel in a neighborhood of
$\boldsymbol{\vartheta}$:
\begin{equation}
d\!\left(
e\!\left(\phi_K(\boldsymbol{\vartheta}')\right)
\right)
=
\phi_K(\boldsymbol{\vartheta}')
\quad
\text{for all $\boldsymbol{\vartheta}'$ in that neighborhood.}
\label{eq:app_exact_ae}
\end{equation}
Then
\begin{equation}
m\geq 4K.
\label{eq:app_latent_lower_bound}
\end{equation}
The physical path tuple uses exactly $4K$ real coordinates and therefore
attains this local lower bound.

\noindent\textit{Proof.}
Differentiate \eqref{eq:app_exact_ae} with respect to
$\boldsymbol{\vartheta}'$ at $\boldsymbol{\vartheta}$. The chain rule gives
\begin{equation}
J_d\,J_e\,J_{\phi_K}
=
J_{\phi_K}.
\label{eq:app_chain_rule}
\end{equation}
By Assumption A.1, the right-hand side has rank $4K$. The left-hand side
factors through an $m$-dimensional latent space and therefore has rank at most
$m$. Hence $4K\leq m$.

Under local identifiability, the parameters in
\eqref{eq:app_path_parameters} themselves form valid local coordinates, and
\eqref{eq:app_path_synthesis} reconstructs the channel exactly. The
$4K$-coordinate path representation therefore achieves equality. Finally,
$\psi=\arcsin(u)$ is a smooth one-to-one change of coordinates inside the
field of view, so reporting $\psi$ instead of $u$ does not change the
coordinate count.
\hfill$\square$

\subsection{Variable Path Count Versus a Fixed-Width Autoencoder}

\noindent\textbf{Corollary A.1 (variable-rate advantage).}
Suppose one fixed-width autoencoder must exactly represent all identifiable
channels with
\[
1\leq K\leq K_{\max}.
\]
Since this family contains an identifiable $K_{\max}$-path subset,
Theorem A.1 implies
\begin{equation}
m_{\mathrm{fixed}}
\geq
4K_{\max}.
\label{eq:app_fixed_width_bound}
\end{equation}
The path message instead uses
\begin{equation}
P_{\mathrm{path}}(\mathbf H)
=
4K(\mathbf H),
\qquad
\mathbb E[P_{\mathrm{path}}]
=
4\,\mathbb E[K].
\label{eq:app_variable_payload}
\end{equation}
Consequently, it uses strictly fewer real values on every channel with
$K<K_{\max}$, and it has strictly smaller average payload whenever
\[
\Pr[K<K_{\max}]>0.
\]
At $K=K_{\max}$, it meets the $4K_{\max}$ lower bound; an ideal autoencoder
with the same latent dimension may tie this bound.

Thus, under the ideal identifiable model, both representations may reconstruct
the channel exactly, but the path representation uses a strictly smaller
message whenever the current channel requires fewer than $K_{\max}$ paths.

\paragraph{Scope.}
The result is deliberately narrow. It applies to differentiable, locally
exact codecs with a fixed real-coordinate latent for the identifiable
geometric component before quantization. It does not claim that every trained
autoencoder must perform worse, or that no alternative $4K$-dimensional
coordinates exist. It also does not cover deliberately lossy codes,
entropy-coded bit strings, coincident paths, or the diffuse and omitted energy
outside the retained $K$-path model. Those effects are measured empirically
by the deployable NMSE in the main paper.

%% file: Sections/Appendix_B.tex
\section{GCNO Details and Multipath Interpretation}
\label{app:gcno_theory}

This section gives the normalization and stability details omitted from the
main text and then connects the Gramian--Chebyshev operation directly to the
geometric multipath model. Bold symbols denote vectors or matrices, and
\[
\langle\mathbf A,\mathbf B\rangle_F
=
\operatorname{tr}
\left(\mathbf A^{\mathsf H}\mathbf B\right).
\]

\subsection{Multipath Footprints in the Evidence Map}
\label{app:evidence_factorization}

Consider first the retained $K$-path component
\begin{equation}
\mathbf H_K
=
\sum_{k=1}^{K}
g_k\,
\mathbf a_{N_r}(u_{r,k})
\mathbf a_{N_t}(u_{t,k})^{\mathsf H}.
\label{eq:app_HK}
\end{equation}
For the receive and transmit grids in the main paper, define the
one-dimensional matched-filter footprints
\begin{align}
\mathbf b_{r,k}[i]
&=
\mathbf a_{N_r}(u_i^r)^{\mathsf H}
\mathbf a_{N_r}(u_{r,k}),
\label{eq:app_br}\\
\mathbf b_{t,k}[j]
&=
\mathbf a_{N_t}(u_j^t)^{\mathsf H}
\mathbf a_{N_t}(u_{t,k}).
\label{eq:app_bt}
\end{align}
For the centered steering convention of the main paper, each entry is a
normalized Dirichlet kernel (with removable singularities defined by continuity),
\begin{equation}
\kappa_N(\delta)
=
\frac{\sin(N\pi\delta/2)}
     {N\sin(\pi\delta/2)},
\qquad
\label{eq:app_dirichlet}
\end{equation}
where $\delta$ is the difference between the physical and grid spatial
coordinates. Thus, an off-grid path produces a broad receive footprint times
a broad transmit footprint.

Let
\[
\mathbf B_r
=
[\mathbf b_{r,1},\ldots,\mathbf b_{r,K}],
\qquad
\mathbf B_t
=
[\mathbf b_{t,1},\ldots,\mathbf b_{t,K}],
\]
and
\[
\mathbf\Gamma
=
\operatorname{diag}(g_1,\ldots,g_K).
\]

\noindent\textbf{Lemma B.1 (exact evidence-map factorization).}
For the matched-filter map
\[
C_{ij}
=
\langle\mathbf D_{ij},\mathbf H_K\rangle_F,
\]
we have
\begin{equation}
\mathbf C
=
\sum_{k=1}^{K}
g_k\mathbf b_{r,k}\mathbf b_{t,k}^{\mathsf H}
=
\mathbf B_r\mathbf\Gamma\mathbf B_t^{\mathsf H},
\qquad
\operatorname{rank}(\mathbf C)\leq K.
\label{eq:app_C_factorization}
\end{equation}

\noindent\textit{Proof.}
Substituting one path into the Frobenius inner product gives
\begin{equation}
\left\langle
\mathbf D_{ij},
g_k\mathbf a_{N_r}(u_{r,k})
\mathbf a_{N_t}(u_{t,k})^{\mathsf H}
\right\rangle_F
=
g_k\,
\mathbf b_{r,k}[i]\,
\mathbf b_{t,k}[j]^*.
\end{equation}
Summing over $k$ gives \eqref{eq:app_C_factorization}. The rank bound follows
because the result is a sum of at most $K$ rank-one matrices.
\hfill$\square$

The two evidence-map Gramians therefore satisfy
\begin{align}
\mathbf C\mathbf C^{\mathsf H}
&=
\mathbf B_r\mathbf\Gamma
(\mathbf B_t^{\mathsf H}\mathbf B_t)
\mathbf\Gamma^{\mathsf H}\mathbf B_r^{\mathsf H},
\label{eq:app_C_gram_r}\\
\mathbf C^{\mathsf H}\mathbf C
&=
\mathbf B_t\mathbf\Gamma^{\mathsf H}
(\mathbf B_r^{\mathsf H}\mathbf B_r)
\mathbf\Gamma\mathbf B_t^{\mathsf H}.
\label{eq:app_C_gram_t}
\end{align}
Hence their active receive and transmit subspaces are contained in
$\operatorname{span}(\mathbf B_r)$ and
$\operatorname{span}(\mathbf B_t)$, respectively. The diagonal contributions
encode individual path strengths, while the off-diagonal terms encode overlap
between nearby path footprints. If the measured channel also contains omitted
or diffuse energy, its matched-filter contribution is simply added to
$\mathbf C$ by linearity.

\subsection{Normalized Gramian--Chebyshev Filtering}
\label{app:gcno_normalization}

For the complex feature state
\[
\left\{
\mathbf X_c^{(\ell)}
\right\}_{c=1}^{C_\ell},
\]
GCNO forms the Gramians given in the main paper and normalizes each side
$s\in\{r,t\}$ as
\begin{equation}
\overline{\mathbf G}_s^{(\ell)}
=
\frac{
\mathbf G_s^{(\ell)}+\epsilon\mathbf I
}{
\operatorname{tr}
\left(\mathbf G_s^{(\ell)}+\epsilon\mathbf I\right)
},
\qquad
\mathbf A_s^{(\ell)}
=
2\overline{\mathbf G}_s^{(\ell)}-\mathbf I.
\label{eq:app_gram_normalization}
\end{equation}
The Chebyshev matrices are evaluated without an eigendecomposition through
\begin{equation}
\footnotesize
T_0(\mathbf A)=\mathbf I,
T_1(\mathbf A)=\mathbf A,
\qquad
T_{q+1}(\mathbf A)
=
2\mathbf A T_q(\mathbf A)-T_{q-1}(\mathbf A).
\label{eq:app_cheb_recurrence}
\end{equation}
The implementation uses $q=0,\ldots,Q$ with $Q=3$.

\noindent\textbf{Proposition B.1
(stable filtering and size-independent weights).}
For every sample and layer, all eigenvalues of
$\mathbf A_r^{(\ell)}$ and $\mathbf A_t^{(\ell)}$ lie in $[-1,1]$.
Therefore,
\begin{equation}
\left\|T_p(\mathbf A_r^{(\ell)})\right\|_2
\leq 1,
\qquad
\left\|T_q(\mathbf A_t^{(\ell)})\right\|_2
\leq 1.
\label{eq:app_cheb_bound}
\end{equation}
For one output channel of the linear GCNO core,
\begin{equation}
\left\|
\mathcal F_\ell(\mathbf X)_c
\right\|_F
\leq
\sum_{c'=1}^{C_\ell}
\sum_{p,q=0}^{Q}
\left|
\Theta^{(\ell)}_{cc'pq}
\right|
\left\|
\mathbf X_{c'}^{(\ell)}
\right\|_F.
\label{eq:app_gcno_stability}
\end{equation}
Moreover, the learned coefficients are indexed by feature channels and
polynomial orders, not by receive or transmit locations.

\noindent\textit{Proof.}
Each Gramian is Hermitian positive semidefinite. Equation
\eqref{eq:app_gram_normalization} therefore places the eigenvalues of
$\overline{\mathbf G}_s^{(\ell)}$ in $[0,1]$ and those of
$\mathbf A_s^{(\ell)}$ in $[-1,1]$. Since
\[
|T_q(\lambda)|\leq 1
\qquad
\text{for every }\lambda\in[-1,1],
\]
the spectral-norm bounds in \eqref{eq:app_cheb_bound} follow. Applying
submultiplicativity and the triangle inequality to the bilateral filter in
the main paper gives \eqref{eq:app_gcno_stability}. The final statement
follows directly from the indices of
$\Theta^{(\ell)}_{cc'pq}$. Array and grid sizes enter through the analytical
projections and sample-specific Gramians, while the learned coefficients
remain unchanged.
\hfill$\square$

\subsection{Neural-Operator Characterization}
\label{app:gcno_operator_characterization}

A neural operator parameterizes a map between function
spaces using learned coefficients that are independent of one
particular sampling resolution. Let
\(
\mathcal H_C
=
L^2(\Omega_r\times\Omega_t;\mathbb C^C)
\)
and let
\(
\mathcal X=(\mathcal X_1,\ldots,\mathcal X_C)
\in\mathcal H_C
\)
denote a complex feature field. Define its receive- and
transmit-side Gramian kernels by
\begin{align}
k_r^{\mathcal X}(u,u')
&=
\sum_{c=1}^{C}
\int_{\Omega_t}
\mathcal X_c(u,v)
\overline{\mathcal X_c(u',v)}
\,dv,\\
k_t^{\mathcal X}(v,v')
&=
\sum_{c=1}^{C}
\int_{\Omega_r}
\overline{\mathcal X_c(u,v)}
\mathcal X_c(u,v')
\,du.
\end{align}
Let $\mathcal G_r[\mathcal X]$ and
$\mathcal G_t[\mathcal X]$ be the corresponding positive
semidefinite integral operators. For
\(
E(\mathcal X)=\sum_c\|\mathcal X_c\|_{L^2}^2>0
\),
define
\begin{equation}
\overline{\mathcal G}_s[\mathcal X]
=
\frac{\mathcal G_s[\mathcal X]}{E(\mathcal X)},
\qquad
\mathcal A_s[\mathcal X]
=
2\overline{\mathcal G}_s[\mathcal X]-\mathcal I_s,
\quad s\in\{r,t\}.
\end{equation}
The function-space GCNO core is
\begin{equation}
\begin{aligned}
[\mathcal F_{\Theta}^{(\ell)}(\mathcal X)]_o
&=
\sum_{c=1}^{C_\ell}
\sum_{p,q=0}^{Q}
\Theta^{(\ell)}_{ocpq}\\
&\quad\times
T_p\!\left(\mathcal A_r[\mathcal X]\right)
\mathcal X_c
T_q\!\left(\mathcal A_t[\mathcal X]\right),
\end{aligned}
\label{eq:app_function_space_gcno}
\end{equation}
where the left and right actions operate on the receive and
transmit coordinates, respectively. Residual addition and
the shared pointwise nonlinearity are then composed with
this core as in the implemented layer.

\noindent\textbf{Operator characterization.}
Equation~\eqref{eq:app_function_space_gcno} defines a
nonlinear map
\(
\mathcal H_{C_\ell}\rightarrow
\mathcal H_{C_{\ell+1}}
\),
because its Gramian operators depend on the current input
field. Under the uniform sampling used here, quadrature
reduces the two kernels to matrices proportional to
\(
\sum_c\mathbf X_c\mathbf X_c^{\mathsf H}
\)
and
\(
\sum_c\mathbf X_c^{\mathsf H}\mathbf X_c
\).
The common quadrature factors cancel under normalization,
giving exactly the bilateral matrix filter, with $\epsilon\mathbf I$ serving only
as finite-dimensional numerical regularization.

The learned tensor
\(
\Theta^{(\ell)}_{ocpq}
\)
is indexed only by feature channels and polynomial orders;
no learned index ranges over a receive position, transmit
position, antenna count, or grid cell. Consequently, the same
learned coefficients define compatible discrete realizations
on different admissible receive--transmit samplings after the
fixed analytical projections and sample Gramians are
recomputed. GCNO therefore satisfies the neural-operator
criterion used in the main paper: it learns a function-space
transformation rather than a dimension-specific matrix map.

This is an architectural transfer property, not a claim of
exact equality across resolutions or arbitrary array
geometries. Approximation quality under a changed
discretization remains an empirical question. The shared
stems, LocalGate corrections, and support head are auxiliary
maps around the Gramian--Chebyshev backbone and introduce
no location-specific learned parameters.

\subsection{Physical Interpretation of the Chebyshev Core}
\label{app:gcno_multipath_action}

For one implemented GCNO layer, define the joint receive and transmit
subspaces of its input feature maps as
\begin{equation}
\scriptsize
\mathcal U_\ell
=
\operatorname{span}\!\left(
\bigcup_{c=1}^{C_\ell}
\operatorname{range}(\mathbf X_c^{(\ell)})
\right),
\mathcal V_\ell
=
\operatorname{span}\!\left(
\bigcup_{c=1}^{C_\ell}
\operatorname{range}
\left((\mathbf X_c^{(\ell)})^{\mathsf H}\right)
\right).
\label{eq:app_joint_subspaces}
\end{equation}

\noindent\textbf{Proposition B.2
(subspace action of the GCNO core).}
For every output channel of the linear bilateral filter in the main paper,
\begin{equation}
\operatorname{range}
\left(\mathcal F_\ell(\mathbf X)_c\right)
\subseteq
\mathcal U_\ell,
\qquad
\operatorname{range}
\left(\mathcal F_\ell(\mathbf X)_c^{\mathsf H}\right)
\subseteq
\mathcal V_\ell.
\label{eq:app_multichannel_subspace}
\end{equation}
Thus, the linear core mixes the receive and transmit modes already present in
the current multi-channel state rather than introducing a fixed external
basis.

For a single feature map
\[
\mathbf X
=
\mathbf U\mathbf\Sigma\mathbf V^{\mathsf H},
\]
let $\mathbf\Lambda_r$ and $\mathbf\Lambda_t$ denote the eigenvalues of the
normalized Gramian operators on the active subspaces. The corresponding
single-map core satisfies
\begin{equation}
\footnotesize
\sum_{p,q=0}^{Q}
\theta_{pq}
T_p(\mathbf A_r)
\mathbf X
T_q(\mathbf A_t)
=
\mathbf U
\left[
\sum_{p,q=0}^{Q}
\theta_{pq}
T_p(\mathbf\Lambda_r)
\mathbf\Sigma
T_q(\mathbf\Lambda_t)
\right]
\mathbf V^{\mathsf H}.
\label{eq:app_svd_action}
\end{equation}
In particular, when $\mathbf X=\mathbf C$, Lemma B.1 places these active
subspaces inside the spans of the physical receive and transmit footprints.

If, in addition, the normalized footprints
\[
\widetilde{\mathbf b}_{r,k}
=
\frac{\mathbf b_{r,k}}{\|\mathbf b_{r,k}\|_2},
\qquad
\widetilde{\mathbf b}_{t,k}
=
\frac{\mathbf b_{t,k}}{\|\mathbf b_{t,k}\|_2}
\]
are mutually orthonormal on each side, then the evidence-map core reduces to
\begin{equation}
\sum_{p,q=0}^{Q}
\theta_{pq}
T_p(\mathbf A_r)
\mathbf C
T_q(\mathbf A_t)
=
\sum_{k=1}^{K}
\widetilde g_k\,
h_{\Theta}(\xi_{r,k},\xi_{t,k})\,
\widetilde{\mathbf b}_{r,k}
\widetilde{\mathbf b}_{t,k}^{\mathsf H},
\label{eq:app_pathwise_action}
\end{equation}
where
\begin{equation}
\widetilde g_k
=
g_k
\|\mathbf b_{r,k}\|_2
\|\mathbf b_{t,k}\|_2,
\qquad
h_{\Theta}(x,y)
=
\sum_{p,q=0}^{Q}
\theta_{pq}T_p(x)T_q(y),
\label{eq:app_transfer_function}
\end{equation}
and
\[
\xi_{r,k},\xi_{t,k}\in[-1,1]
\]
are the normalized Gramian eigenvalues associated with path $k$. In this
separated-path case, the core therefore applies a learned bivariate polynomial
response to each physical path footprint.

\noindent\textit{Proof.}
The subspace $\mathcal U_\ell$ is invariant under
$\mathbf G_r^{(\ell)}$ because each term
\[
\mathbf X_c^{(\ell)}
(\mathbf X_c^{(\ell)})^{\mathsf H}
\]
maps into
\[
\operatorname{range}(\mathbf X_c^{(\ell)})
\subseteq
\mathcal U_\ell.
\]
It is consequently invariant under $\mathbf A_r^{(\ell)}$ and every polynomial
of that operator. The same argument applied to the adjoint feature maps shows
that $\mathcal V_\ell$ is invariant under every polynomial of
$\mathbf A_t^{(\ell)}$. Left and right polynomial filtering, channel mixing,
and summation therefore give \eqref{eq:app_multichannel_subspace}.

For the single-map case, $\mathbf A_r$ and $\mathbf A_t$ are affine functions
of
\[
\mathbf X\mathbf X^{\mathsf H}
\qquad\text{and}\qquad
\mathbf X^{\mathsf H}\mathbf X,
\]
so they share the active eigenvectors $\mathbf U$ and $\mathbf V$. A matrix
polynomial preserves these eigenvectors; substitution gives
\eqref{eq:app_svd_action}. Under footprint orthogonality, each normalized
footprint is itself a Gramian eigenvector, and applying the same identity path
by path yields \eqref{eq:app_pathwise_action}.
\hfill$\square$

\paragraph{Interpretation and scope.}
Lemma B.1 shows that a $K$-path channel produces at most $K$ separable
matched-filter footprints. Proposition B.2 shows that the implemented linear
GCNO core filters the joint receive/transmit modes of its current features;
for the ideal evidence map, these modes lie in the physical footprint spans.
When the footprints are separated, the action becomes the explicit path-wise
response $h_{\Theta}$. When footprints overlap, the Gramian modes can be
mixtures of nearby paths, so the correct interpretation is filtering of their
joint physical subspace rather than independent processing of each path.

The full GCNO also contains pointwise nonlinearities and local sharpening.
We therefore make no claim that every hidden feature remains rank $K$; the
result characterizes the physics-aligned linear Gramian--Chebyshev core.

\subsection{Complete Processing Pipeline}
\label{app:complete_pipeline}

The preceding subsections characterized the evidence-map structure and the
Gramian--Chebyshev core. We now give the complete implemented path from an
observed channel to the transmitted tuple set. The detailed record below refers
to the ASU reference model; Dallas and Seattle use the same method and are
trained separately on their corresponding splits.

For the reference configuration, $N_r=N_t=32$ and the receive and transmit
direction grids each contain $R=28$ uniformly spaced spatial coordinates over
$[-\sin(75^\circ),\sin(75^\circ)]$. The grid spacing is
\[
\delta_u = 0.07155.
\]
Every input is first normalized as
\begin{equation}
h_{\mathrm{scale}}=\lVert H\rVert_F,
\qquad
H_{\mathrm{norm}}
=
\frac{H}{h_{\mathrm{scale}}}.
\label{eq:app_channel_normalization}
\end{equation}
Only $H_{\mathrm{norm}}$ is used by the neural feature extractor. The scale
does not require a separate transmitted field: solving for gains using
$H_{\mathrm{norm}}$ and multiplying the fitted gains by
$h_{\mathrm{scale}}$ is equivalent to fitting the gains directly to $H$.

The fixed matched-filter evidence is
\begin{equation}
C_{ij}
=
\left\langle D_{ij},H_{\mathrm{norm}}\right\rangle_F,
\qquad
C\in\mathbb C^{28\times28},
\label{eq:app_evidence_implementation}
\end{equation}
where $D$, $D^r$, and $D^t$ are analytical tensors and are excluded from the
optimizer. 

For the centered steering convention, let
$p_{N,n}=n-(N-1)/2$. The steering derivative with respect to
the spatial coordinate is
\begin{equation}
    \mathbf a'_N(u)[n]
    =
    j\pi p_{N,n}\mathbf a_N(u)[n].
\end{equation}
Accordingly, the fixed derivative dictionaries are
\begin{equation}
\begin{aligned}
    \mathbf D^r_{ij}
    &=
    \mathbf a'_{N_r}(u_i^r)
    \mathbf a_{N_t}(u_j^t)^{\mathsf H},\\
    \mathbf D^t_{ij}
    &=
    \mathbf a_{N_r}(u_i^r)
    \mathbf a'_{N_t}(u_j^t)^{\mathsf H}.
\end{aligned}
\end{equation}
These tensors are constructed analytically once for each array
configuration and remain non-trainable.

For offsets $|\delta_r|,|\delta_t|\leq\delta_u/2$, define the exact
and first-order atoms
\begin{equation}
\begin{aligned}
    \mathbf A_{ij}(\delta_r,\delta_t)
    &=
    \mathbf a_{N_r}(u_i^r+\delta_r)
    \mathbf a_{N_t}(u_j^t+\delta_t)^{\mathsf H},\\
    \mathbf B^{(1)}_{ij}(\delta_r,\delta_t)
    &=
    \mathbf D_{ij}
    +\delta_r\mathbf D^r_{ij}
    +\delta_t\mathbf D^t_{ij}.
\end{aligned}
\end{equation}
The best scalar multiple of the unrefined grid atom has normalized
squared error
\begin{equation}
    \epsilon_{\mathrm{grid}}
    =
    1-
    |\kappa_{N_r}(\delta_r)|^2
    |\kappa_{N_t}(\delta_t)|^2,
\end{equation}
where $\kappa_N$ is defined in Eq.~(13). By first-order expansion,
\begin{equation}
    \left\|
        \mathbf A_{ij}(\delta_r,\delta_t)
        -\mathbf B^{(1)}_{ij}(\delta_r,\delta_t)
    \right\|_F
    =
    \mathcal O\!\left(
        \delta_r^2+\delta_t^2+|\delta_r\delta_t|
    \right),
\end{equation}
and hence its normalized squared approximation error is
\begin{equation}
    \epsilon_{\mathrm{Taylor}}
    =
    \mathcal O\!\left[
        \left(
            \delta_r^2+\delta_t^2+|\delta_r\delta_t|
        \right)^2
    \right].
\end{equation}
The $|\delta_r\delta_t|$ term is the omitted receive--transmit
cross term; the remaining omitted terms are second order along
the individual direction axes.

The channel and evidence are processed by separate lightweight
branches. The channel branch preserves the antenna-domain observation before
projecting its learned features to the direction grid, whereas the evidence
branch directly processes the grid-domain matched-filter map. Their outputs
are fused and passed through the three GCNO/LocalGate layers described below.

Table~\ref{tab:app_tensor_flow} gives the complete tensor flow. Complex
quantities are stored in \texttt{complex64}; neural real-valued channels and
scalar metadata use \texttt{float32}. ``Fixed'' denotes an analytical
operation whose coefficients are not trainable. Such a fixed linear operation
may still pass gradients to its learned input features.

\begin{table*}[t]
\centering
\small
\caption{End-to-end tensor flow for GCNO.}
\label{tab:app_tensor_flow}
\begin{tabular}{l|l|l|p{0.34\textwidth}|l}
\hline
Stage & Output shape & Type & Operation & Status \\
\hline
Observed channel
& $N_b\times32\times32$
& complex
& Physical channel $H$
& Input \\

Channel scale
& $N_b$
& real
& $h_{\mathrm{scale}}=\lVert H\rVert_F$
& Fixed \\

Normalized channel
& $N_b\times32\times32$
& complex
& $H/h_{\mathrm{scale}}$
& Fixed \\

Channel-branch input
& $N_b\times2\times32\times32$
& real
& Real and imaginary parts of $H_{\mathrm{norm}}$
& Fixed \\

Evidence map
& $N_b\times28\times28$
& complex
& $C_{ij}=\langle D_{ij},H_{\mathrm{norm}}\rangle_F$
& Fixed \\

Evidence-branch input
& $N_b\times2\times28\times28$
& real
& Real and imaginary parts of $C$
& Fixed \\

Channel stem
& $N_b\times24\times32\times32$
& complex
& $2\!\rightarrow\!96\!\rightarrow\!48$ real channels, paired into
24 complex channels
& Learned \\

Channel-grid projection
& $N_b\times24\times28\times28$
& complex
& Projection of every learned channel feature through fixed $D$
& Fixed \\

Evidence stem
& $N_b\times24\times28\times28$
& complex
& $2\!\rightarrow\!96\!\rightarrow\!48$ real channels, paired into
24 complex channels
& Learned \\

Fusion input
& $N_b\times98\times28\times28$
& real
& Real and imaginary parts of both 24-channel branches, plus two
coordinate maps
& Mixed \\

Fused state
& $N_b\times24\times28\times28$
& complex
& $98\!\rightarrow\!96\!\rightarrow\!48$ real channels, paired into
24 complex channels
& Learned \\

GCNO states
& $N_b\times24\times28\times28$
& complex
& Three bilateral Gramian--Chebyshev layers, each followed by a
LocalGate
& Mixed \\

Head features
& $N_b\times73\times28\times28$
& real
& $\Re X$, $\Im X$, $|X|$, and
$\log\!\left(\sum_c|X_c|^2\right)$
& Fixed \\

Head outputs
& $N_b\times3\times28\times28$
& real
& One score map and two bounded Taylor-offset maps
& Learned \\

Selected atoms
& $N_b\times K\times32\times32$
& complex
& Taylor atoms during training; exact steering atoms during
deployment
& Fixed \\

Fitted gains
& $N_b\times K$
& complex
& Joint ridge least squares
& Fixed \\

Transmitted message
& $N_b\times4K$
& real
& $(\Re\widehat g_k,\Im\widehat g_k,
\widehat\psi_{r,k},\widehat\psi_{t,k})_{k=1}^{K}$
& Output \\
\hline
\end{tabular}
\end{table*}

The learned network therefore predicts only support priorities and local
direction corrections. It does not predict gains and does not reconstruct the
channel through a neural decoder. During training, a smooth selector and
Taylor atoms make the tuple-based reconstruction differentiable. During
deployment, the smooth approximation is discarded: the UE performs adaptive
hard support selection, generates exact analytical atoms under no gradient,
and jointly fits the gains. The BS receives only the resulting tuples,
converts $\widehat\psi$ to $\widehat u=\sin(\widehat\psi)$, and evaluates the
known geometric synthesis model.

\subsection{Layer-by-Layer GCNO Architecture}
\label{app:layerwise_architecture}

\paragraph{Dual input branches.}
The channel and evidence branches use identical stem widths but do not share
parameters. Each stem applies
\[
\operatorname{Conv}_{1\times1}(2,96)
\;\rightarrow\;
\operatorname{GELU}
\;\rightarrow\;
\operatorname{Conv}_{1\times1}(96,48).
\]
The 48 real output channels are interpreted as the real and imaginary parts
of 24 complex feature maps. The channel-branch maps are initially sampled on
the $32\times32$ antenna grid and are projected to the $28\times28$
direction grid using the fixed dictionary. The evidence branch already
operates on the direction grid.

The real and imaginary parts of both branch outputs provide 96 real channels.
Two normalized receive/transmit coordinate maps are appended, producing the
98-channel fusion input. The fusion stem applies
\[
\operatorname{Conv}_{1\times1}(98,96)
\;\rightarrow\;
\operatorname{GELU}
\;\rightarrow\;
\operatorname{Conv}_{1\times1}(96,48),
\]
and again pairs the result into a 24-channel complex state.

\paragraph{Bilateral GCNO layers.}
The fused state passes through exactly three GCNO layers. Each layer uses the
normalized receive and transmit Gramian operators defined in
Sec.~\ref{app:complete_pipeline} and the preceding theoretical subsections.
The Chebyshev order is three, so each side contains the four basis matrices
$T_0,T_1,T_2,T_3$. For 24 input and 24 output complex channels, one layer has
the coefficient tensor
\[
\Theta^{(\ell)}
\in
\mathbb C^{24\times24\times4\times4}.
\]
All 16 receive--transmit polynomial pairs are applied:
\begin{equation}
Y^{(\ell)}_o
=
\sum_{c=1}^{24}
\sum_{p=0}^{3}
\sum_{q=0}^{3}
\Theta^{(\ell)}_{o c p q}
T_p\!\left(A_r^{(\ell)}\right)
X_c^{(\ell)}
T_q\!\left(A_t^{(\ell)}\right).
\label{eq:app_layerwise_gcno}
\end{equation}
The layer adds the input state as a residual connection and applies GELU
separately to the real and imaginary parts.

\paragraph{LocalGate.}
A LocalGate follows each of the three GCNO layers. It has two complementary
paths. The local path converts the 24 complex maps into 48 real channels,
applies a $3\times3$ depthwise convolution, GELU, and a $1\times1$
projection, and returns the result through a learned residual scale. This path
provides local sharpening without replacing the global Gramian operation.
The channel path averages the power of each complex feature channel and
applies a $24\!\rightarrow\!12\!\rightarrow\!24$ MLP with GELU and sigmoid.
Its multiplicative response is centered at one, so it softly rescales rather
than removes feature channels.

\paragraph{Support and offset head.}
After the third LocalGate, the complex state is represented by
\[
\left[
\Re X,\;
\Im X,\;
|X|,\;
\log\!\left(\sum_{c=1}^{24}|X_c|^2\right)
\right],
\]
which contains $24+24+24+1=73$ real channels. The compact head is
\[
\begin{aligned}
73
&\xrightarrow{\;1\times1\ \mathrm{Conv}\;} 64
\xrightarrow{\;\mathrm{GELU}\;} 64 \\
&\xrightarrow{\;3\times3\ \mathrm{DWConv}\;} 64
\xrightarrow{\;\mathrm{GELU}\;} 64
\xrightarrow{\;1\times1\ \mathrm{Conv}\;} 3 .
\end{aligned}
\]
Its first output is the unconstrained score map $S$. The remaining outputs are
converted to receive and transmit corrections by
\begin{equation}
\Delta U^r
=
\frac{\delta_u}{2}\tanh Z_r,
\qquad
\Delta U^t
=
\frac{\delta_u}{2}\tanh Z_t.
\label{eq:app_offset_bound}
\end{equation}
Thus each correction remains within one half-cell of its anchor. No gain head
is present.

\paragraph{Exact trainable parameter count.}
Table~\ref{tab:app_parameter_count} counts real scalar parameters. A complex
coefficient contributes two real scalars. Fixed dictionaries, coordinate
maps, Taylor construction, support selection, and LS contain no trainable
parameters.

\begin{table}[t]
\centering
\small
\caption{Trainable parameters of the implemented GCNO model.}
\label{tab:app_parameter_count}
\begin{tabular}{l|r}
\hline
Component & Real parameters \\
\hline
Channel stem, $2\!\rightarrow\!96\!\rightarrow\!48$ & 4,944 \\
Evidence stem, $2\!\rightarrow\!96\!\rightarrow\!48$ & 4,944 \\
Fusion stem, $98\!\rightarrow\!96\!\rightarrow\!48$ & 14,160 \\
GCNO layer 1 & 18,432 \\
GCNO layer 2 & 18,432 \\
GCNO layer 3 & 18,432 \\
Three local residual paths & 8,499 \\
Three channel gates & 1,839 \\
Score/offset head & 5,571 \\
\hline
\textbf{Total} & \textbf{95,253} \\
\hline
\end{tabular}
\end{table}

The checkpoint contains 54 parameter tensors and 95,253 trainable real
scalars, which is the $95$K model size reported in the main paper.

\subsection{Training Objective and Optimization}
\label{app:training_details}

\paragraph{Strictly label-free optimization.}
Training begins from random initialization and does not use a pretrained
checkpoint, teacher warmup, signal-processing pseudolabel, path label,
oracle path count, oracle gain, oracle direction, or external warm start.
Every differentiable quantity is computed from $H_{\mathrm{norm}}$, the fixed
dictionaries, model outputs, or the corresponding LS reconstruction. The same
restriction applies to validation, checkpoint selection, the controller, and
operating-point selection.

\paragraph{Smooth candidate selection.}
Training exposes at most $M=8$ soft candidates. Let
$\mathcal W_m$ denote the current selector window for candidate $m$. A
temperature-controlled distribution over that window is
\begin{equation}
p_m(i,j)
=
\frac{
\exp\!\left(S_m(i,j)/\tau\right)
}{
\sum_{(a,b)\in\mathcal W_m}
\exp\!\left(S_m(a,b)/\tau\right)
},
\qquad
(i,j)\in\mathcal W_m.
\label{eq:app_soft_assignment}
\end{equation}
The corrected continuous center is the probability-weighted expectation
\begin{equation}
\boldsymbol\mu_m
=
\sum_{(i,j)\in\mathcal W_m}
p_m(i,j)
\begin{bmatrix}
u_i^r+\Delta U^r_{ij}\\
u_j^t+\Delta U^t_{ij}
\end{bmatrix}.
\label{eq:app_soft_center}
\end{equation}
After each candidate, a Gaussian penalty with width $0.8$ grid cells and
logit penalty $8.5$ suppresses its neighborhood before the next soft
candidate is formed. This encourages different slots to cover distinct
regions. A sigmoid activity
$\alpha_m\in(0,1)$ controls the contribution of candidate $m$, and
\[
\widehat K_{\mathrm{soft}}
=
\sum_{m=1}^{8}\alpha_m
\]
is the differentiable estimate of the retained path count.

For the anchor $(i_m,j_m)$ associated with candidate $m$, the training atom is
\begin{equation}
B_m
=
D_{i_mj_m}
+
\Delta u_{r,m}D^r_{i_mj_m}
+
\Delta u_{t,m}D^t_{i_mj_m}.
\label{eq:app_training_taylor_atom}
\end{equation}
Exact steering vectors are not generated inside this differentiable path.

Let
\[
h=\operatorname{vec}(H_{\mathrm{norm}}),
\qquad
\mathcal B
=
\left[
\operatorname{vec}(B_1),\ldots,
\operatorname{vec}(B_8)
\right].
\]
Differentiable complex ridge LS uses $\eta_{\mathrm{train}}=10^{-4}$:
\begin{equation}
\widehat{\boldsymbol g}
=
\left(
\mathcal B^{\mathrm H}\mathcal B
+
\eta_{\mathrm{train}}I
\right)^{-1}
\mathcal B^{\mathrm H}h,
\qquad
\widehat h_{\mathrm{soft}}
=
\mathcal B
\left(
\boldsymbol\alpha\odot\widehat{\boldsymbol g}
\right).
\label{eq:app_training_ls}
\end{equation}

\paragraph{Objective.}
The complete one-stage objective used by the reference model is
\begin{align}
\mathcal L
={}&
\mathbb E_H
\left[
\log\left(
\frac{
\lVert h-\widehat h_{\mathrm{soft}}\rVert_2^2
}{
\lVert h\rVert_2^2+\epsilon
}
+\epsilon
\right)
\right]
+
0.04\,\mathcal L_{\mathrm{rate}}
\nonumber\\
&+
0.02\,\mathcal L_{\mathrm{dup}}
+
10^{-4}\mathcal L_{\mathrm{off}}
+
10^{-4}\mathcal L_{\mathrm{score}}.
\label{eq:app_complete_training_loss}
\end{align}
Here
\[
\mathcal L_{\mathrm{rate}}
=
\mathbb E_H
\left[
\sum_{m=1}^{8}\alpha_m
\right].
\]
The duplicate term $\mathcal L_{\mathrm{dup}}$ is the activity-weighted
coherence penalty between candidate atoms. The offset term
$\mathcal L_{\mathrm{off}}$ penalizes receive and transmit corrections after
normalization by the half-cell bound $\delta_u/2$. The score term is the
spatial mean of the sigmoid-activated score map,
\begin{equation}
\mathcal L_{\mathrm{score}}
=
\mathbb E_H
\left[
\frac{1}{R^2}
\sum_{i,j}\sigma(S_{ij})
\right].
\label{eq:app_score_penalty}
\end{equation}
These terms respectively discourage unnecessary paths, repeated atoms,
cell-edge corrections, and broad score activation. No label-derived
quantity appears in any term.

\paragraph{Initialization and optimizer.}
The exact optimization settings are summarized in
Table~\ref{tab:app_optimizer}. Complex Chebyshev coefficients are initialized
from a zero-mean normal distribution with scale
\[
\frac{0.05}{\sqrt{24\cdot16}}.
\]
The offset-output weights and biases are initialized to zero. The final
projections in the local and channel gates use a zero-mean normal
initialization of scale $10^{-3}$ with zero bias. The score-output bias is
initialized to $-1.45$.

\begin{table}[t]
\centering
\small
\caption{Optimization settings for the ASU reference run.}
\label{tab:app_optimizer}
\begin{tabular}{l|l}
\hline
Setting & Value \\
\hline
Initialization & Random; no checkpoint \\
Optimizer & Adam \\
Batch size & 128 \\
Adam $(\beta_1,\beta_2)$ & $(0.9,0.999)$ \\
Adam $\epsilon$ & $10^{-8}$ \\
Weight decay & $10^{-6}$ \\
Global/local learning rate & $3{\times}10^{-4}$ / $10^{-4}$ \\
Gradient clipping & Global norm $5$ \\
Mixed precision & Disabled \\
Matrix-multiplication precision & High \\
Training ridge & $10^{-4}$ \\
Maximum soft candidates & 8 \\
Configured epoch ceiling & 78 \\
Realized controller trajectory & 71 epochs \\
Selected paper checkpoint & Epoch 42 \\
\hline
\end{tabular}
\end{table}

\paragraph{Localization curriculum.}
Training begins with global candidate selection and progressively narrows the
selector window. The realized 71-epoch controller trajectory is shown in
Table~\ref{tab:app_training_schedule}. Learning-rate changes are controlled by
this phase schedule rather than by any label-dependent scheduler.

\begin{table}[t]
\centering
\small
\caption{Realized selector-window curriculum.}
\label{tab:app_training_schedule}
\begin{tabular}{l|c|c|c|c}
\hline
Phase & Window & Epochs & LR & $\tau$ \\
\hline
Global & $28\times28$ & 30 & $3{\times}10^{-4}$ & 0.035 \\
Local 1 & 21 & 6 & $10^{-4}$ & 0.08 \\
Local 2 & 15 & 6 & $10^{-4}$ & 0.08 \\
Local 3 & 9 & 7 & $10^{-4}$ & 0.08 \\
Local 4 & 5 & 14 & $10^{-4}$ & 0.08 \\
Local 5 & 3 & 8 & $10^{-4}$ & 0.08 \\
\hline
\end{tabular}
\end{table}

The controller uses a relative-improvement threshold of $0.005$ and patience
three, subject to the listed phase durations. The frozen paper checkpoint is
the validation-selected epoch-42 state obtained in the 15-cell phase. The
remaining controller epochs were executed but did not replace that
validation-best state.

Checkpoint and operating-point choices are made from label-free validation
reconstruction and rate quantities. Test channels, ground-truth directions,
gains, physical path counts, oracle reconstructions, angle errors, and
path-matching scores do not select the model or any operating threshold.

\subsection{Adaptive Deployment and LS Reconstruction}
\label{app:hard_deployment}

Deployment replaces the smooth training approximation with a deterministic
UE-side procedure under \texttt{no\_grad}. Let the corrected coordinate
associated with grid cell $(i,j)$ be
\[
\widetilde u^r_{ij}=u_i^r+\Delta U^r_{ij},
\qquad
\widetilde u^t_{ij}=u_j^t+\Delta U^t_{ij}.
\]
The exact continuous atom used at deployment is
\begin{equation}
A(\widetilde u^r,\widetilde u^t)
=
a_{N_r}(\widetilde u^r)
a_{N_t}(\widetilde u^t)^{\mathrm H}.
\label{eq:app_exact_deployment_atom}
\end{equation}
This exact atom is used only outside gradient flow.

For each channel, the hard selector performs the following operations.

\begin{enumerate}
\item Sort all grid candidates in descending order of the GCNO score.
\item Apply the learned Taylor correction to each tested anchor.
\item Reject a candidate whose corrected pattern is numerically
indistinguishable from an already admitted pattern.
\item Generate the candidate's exact analytical steering atom.
\item Temporarily append the atom to the admitted set and jointly refit all
complex gains by ridge LS.
\item Admit the candidate only when it produces the validation-selected
normalized residual reduction and satisfies the numerical-conditioning
safeguard.
\item Stop when no further meaningful reduction is obtained or the
operating-point-specific cap is reached.
\item Jointly refit all final gains and form the transmitted tuple set.
\end{enumerate}

For an admitted set $\mathcal A$, let
\[
\mathbf A_{\mathcal A}
=
\left[
\operatorname{vec}(A_1),\ldots,
\operatorname{vec}(A_{|\mathcal A|})
\right].
\]
The deployment fit uses
\begin{equation}
\widehat{\boldsymbol g}_{\mathcal A}
=
\left(
\mathbf A_{\mathcal A}^{\mathrm H}\mathbf A_{\mathcal A}
+
\eta_{\mathrm{dep}}I
\right)^{-1}
\mathbf A_{\mathcal A}^{\mathrm H}\operatorname{vec}(H),
\qquad
\eta_{\mathrm{dep}}=3{\times}10^{-5}.
\label{eq:app_deployment_ls}
\end{equation}
If $\widehat h_{\mathcal A}$ and
$\widehat h_{\mathcal A\cup\{c\}}$ are the trial reconstructions before and
after adding candidate $c$, its normalized improvement is
\begin{equation}
\Delta(c\mid\mathcal A)
=
\frac{
\lVert h-\widehat h_{\mathcal A}\rVert_2^2
-
\lVert h-\widehat h_{\mathcal A\cup\{c\}}\rVert_2^2
}{
\lVert h\rVert_2^2+\epsilon
}.
\label{eq:app_admission_reduction}
\end{equation}
The candidate is retained only when this quantity exceeds the selected
threshold. The LS-system condition number is capped at $10^4$, and all
reported ASU profiles enforce $K_{\min}=1$.

After support admission, three optional local polishing rounds examine a
$3\times3$ neighborhood around each coordinate using step fractions
$1/2$, $1/4$, and $1/8$. A non-worsening safeguard retains a proposed
coordinate change only when it does not increase the jointly refitted
residual. Polishing changes neither $K$ nor the payload length.

Table~\ref{tab:app_operating_points} gives the validation-locked ASU
operating profiles. The payload columns are the measured adaptive $4K$
distributions, not $4K_{\max}$.

\begin{table}[t]
\centering
\small
\caption{Adaptive support profiles and realized unquantized payloads.}
\label{tab:app_operating_points}
\begin{tabular}{c|c|c|c|c|c}
\hline
Target Region
& $K_{\max}$
& $\Delta_{\min}$
& Mean
& Median \\
\hline
$-12$ dB & 1 & Single candidate & 4.000 & 4 \\
$-14$ dB & 2 & 0.012  & 5.795  & 4  \\
$-16$ dB & 6 & 0.002  & 9.527  & 8  \\
$-18$ dB & 6 & 0.0005 & 13.889 & 12 \\
$-20$ dB & 7 & 0.0005 & 14.372 & 12 \\
\hline
\end{tabular}
\end{table}

The final UE message is
\[
\left\{
\left(
\Re\widehat g_k,
\Im\widehat g_k,
\widehat\psi_{r,k},
\widehat\psi_{t,k}
\right)
\right\}_{k=1}^{K}.
\]
The BS does not run GCNO, repeat support selection, or refit the gains. It
only dequantizes the tuple fields when necessary, computes
$\widehat u=\sin(\widehat\psi)$, and analytically synthesizes the reconstructed
channel.

\subsection{Quantization and Bit Accounting}
\label{app:quantization_details}

Float-domain and bit-domain results are treated as separate operating points.
The unquantized rate is the measured mean number of transmitted real values,
$4\mathbb E[K]$. The quantized rate is the complete packet length after
encoding the selected path count and every retained tuple field. It is not
obtained by multiplying an unquantized payload by a fixed number of bits.

The quantized packet uses a fixed three-bit header for $K$. For a profile with
field allocations
$b_{\Re g}$, $b_{\Im g}$, $b_{\psi_r}$, and $b_{\psi_t}$,
the packet length for one sample is
\begin{equation}
B_{\mathrm{packet}}
=
3
+
\sum_{k=1}^{K}
\left(
b_{\Re g,k}
+
b_{\Im g,k}
+
b_{\psi_r,k}
+
b_{\psi_t,k}
\right).
\label{eq:app_packet_length}
\end{equation}
The scalar codebooks and profile-specific bit allocations are fitted on the
validation set. The profile, codebooks, support threshold, and all associated
ranges are then frozen before the test set is evaluated. No path annotation
or test result is used to construct the quantizer.

For comparison, dense storage of a $32\times32$ complex channel using
32-bit real and imaginary components requires
\begin{equation}
B_{\mathrm{dense}}
=
2(32)(32)(32)
=
65{,}536
\quad\text{bits}.
\label{eq:app_dense_bit_cost}
\end{equation}
The bit-domain compression ratio is consequently
\begin{equation}
\mathrm{CR}_{\mathrm{bits}}
=
\frac{65{,}536}
{\mathbb E[B_{\mathrm{packet}}]}.
\label{eq:app_bit_compression_ratio}
\end{equation}

Because $K$ is sample-dependent, the actual average packet length need not
equal the nominal requested profile. The locked ASU packet means are reported
in Table~\ref{tab:app_quantized_profiles}.

\begin{table}[t]
\centering
\small
\caption{Nominal quantized profiles and measured mean packet lengths on ASU.}
\label{tab:app_quantized_profiles}
\begin{tabular}{c|c}
\hline
Requested profile & Mean transmitted bits \\
\hline
48  & 52.359 \\
64  & 63.354 \\
96  & 82.109 \\
128 & 118.987 \\
\hline
\end{tabular}
\end{table}

Only the quantized tuple fields and the path-count header are transmitted.
Intermediate GCNO features, score maps, activity values, dictionary
coefficients, and neural parameters are not part of the packet. After
dequantization, the BS performs the same analytical reconstruction as in the
unquantized setting.

\subsection{Reproducibility and Computational Cost}
\label{app:reproducibility_compute}

\paragraph{Dataset and splits.}
The detailed reference record uses the DeepMIMO ASU Campus 3.5\,GHz scenario
with $32\times32$ half-wavelength arrays and a $75^\circ$ field of view.
Channel generation uses 512 OFDM subcarriers over 10\,MHz; the center
subcarrier is retained for the narrowband experiment. The fixed disjoint
split contains 10,000 training, 2,000 validation, and 1,500 held-out test
channels. Dallas and Seattle use the same split sizes and are trained separately.

The training data interface exposes the normalized channel and scale required
by the method. Path directions, gains, masks, physical path counts, oracle
values, and path-matching quantities are not loaded by the training
objective. The held-out test set remains unused until the architecture,
checkpoint, quantizer, and operating profiles have been fixed using
validation-only quantities.

\paragraph{Frozen-model environment.}
Table~\ref{tab:app_reference_environment} gives the verified environment of
the frozen ASU model. The reported parameter count includes only
trainable model tensors.

\begin{table}[t]
\centering
\small
\caption{Frozen ASU model and software environment.}
\label{tab:app_reference_environment}
\begin{tabular}{l|l}
\hline
Item & Verified value \\
\hline
GPU & NVIDIA H100, 80\,GB HBM3 \\
Python & 3.11.13 \\
PyTorch & 2.11.0+\texttt{cu128} \\
CUDA runtime & 12.8 \\
NumPy & 1.26.4 \\
PyYAML & 6.0.3 \\
Matplotlib & 3.10.8 \\
Model-state tensors & 54 \\
Trainable parameters & 95,253 \\
\hline
\end{tabular}
\end{table}

\paragraph{Isolated timing reproduction.}
Runtime and memory were measured in a separate compute-only reproduction on
an otherwise idle NVIDIA H100 with 80\,GB HBM3. This benchmark used Python
3.12.8, PyTorch 2.12.0+\texttt{cu130}, CUDA 13.0, and NumPy 2.2.6, while
preserving the exact ASU $R=28$ configuration. 

The benchmark began from random initialization, used batch size 128 with AMP
disabled, and executed the frozen 71-epoch controller trajectory. Ten
disposable warmup optimizer steps initialized the CUDA kernels on a separate
model instance. The synchronized full-fit measurement then processed 5,609
optimizer updates, corresponding to 710,000 sample presentations. The
measurement-only final state was not used to replace or select the paper
checkpoint.

Table~\ref{tab:app_compute_profile} gives the principal compute and memory
measurements. Training wall time includes the required in-loop soft
validation and the package's normal checkpoint writes. The inference-memory
reserved value includes CUDA allocator cache remaining from the preceding
training call; allocated and incremental peaks better represent the live
inference footprint.

\begin{table*}[t]
\centering
\small
\caption{Isolated H100 compute and memory profile.}
\label{tab:app_compute_profile}
\begin{tabular}{l|l|p{0.46\textwidth}}
\hline
Quantity & Measurement & Scope \\
\hline
Full training wall time
& 178.441\,s
& 71 epochs, 5,609 updates, including in-loop validation and normal
checkpoint writes \\

Full-fit throughput
& 3,978.9 samples/s
& Complete synchronized training call \\

Warmed training step
& 27.212\,ms mean
& Batch 128; matched filter, network, soft selection, LS loss, backward
pass, clipping, and Adam update \\

Warmed-step throughput
& 4,703.8 samples/s
& 100 CUDA-event measurements after 10 warmups \\

Training CUDA peak
& 3.173\,GB allocated
& Maximum live allocation during the full training call \\

Inference CUDA peak
& 0.339\,GB allocated
& Maximum across network, adaptive selector, exact steering, LS, and
complete-encoder scopes \\

Inference incremental peak
& 0.226\,GB
& Increase above the resident allocation during inference \\

Maximum process CPU RSS
& 2.262\,GB
& Maximum observed host memory for the benchmark process \\

Parameter storage
& 381,012 bytes
& 95,253 float32 trainable parameters \\

Serialized model state
& 396,865 bytes
& Model state dictionary without optimizer state \\

Paper checkpoint storage
& 1,216,541 bytes
& Serialized training checkpoint \\

GCNO network compute
& 334.285M MACs / 668.569M FLOPs
& One neural forward with precomputed evidence, including the fixed
channel-branch projection \\

Matched-filter preprocessing
& 3.211M MACs / 6.423M FLOPs
& One $28\times28$ complex evidence map from a $32\times32$ channel \\
\hline
\end{tabular}
\end{table*}

Complete UE-encoder latency includes Frobenius normalization, matched-filter
construction, GCNO inference, adaptive support selection, exact steering
generation, joint LS, and all three coordinate-polishing rounds. Batch-1
measurements use 50 warmups and 500 timed calls. Batch-128 throughput uses
10 warmups and 100 timed batches. Disk input/output, host-to-device transfer,
quantization, plotting, validation selection, and external metric aggregation
are excluded.

\begin{table*}[t]
\centering
\small
\caption{End-to-end UE-encoder latency across bundled adaptive operating
profiles. Payload entries are measured $4K$ distributions.}
\label{tab:app_inference_latency}
\begin{tabular}{l|c|c|c|c|c|c}
\hline
Profile
& Payload mean/med./P95
& Mean
& Median
& P95
& P99
& Batch-128 throughput \\
\hline
Low
& 4.000 / 4 / 4
& 10.131\,ms
& 9.930\,ms
& 10.666\,ms
& 15.341\,ms
& 7,996.6 samples/s \\

Mid
& 8.094 / 8 / 16
& 15.425\,ms
& 15.027\,ms
& 20.803\,ms
& 26.702\,ms
& 3,042.3 samples/s \\

High variant
& 15.297 / 16 / 32
& 20.690\,ms
& 20.317\,ms
& 31.973\,ms
& 32.480\,ms
& 1,098.3 samples/s \\

High
& 15.297 / 16 / 32
& 23.830\,ms
& 23.390\,ms
& 35.908\,ms
& 36.773\,ms
& 1,055.2 samples/s \\

Maximum
& 31.922 / 32 / 32
& 32.789\,ms
& 32.183\,ms
& 35.053\,ms
& 44.715\,ms
& 1,528.4 samples/s \\
\hline
\end{tabular}
\end{table*}

The analytical operation count treats one real multiply--accumulate as one
MAC and two FLOPs, and one complex multiply--accumulate as four real MACs and
eight FLOPs. Nonlinearities, exponentials, divisions, square roots, sorting,
linear solves, and condition-number evaluation are not included in these
static MAC/FLOP totals. Because the adaptive selector has data-dependent
control flow, partial profiler attribution is not presented as a complete
decoder FLOP count; measured end-to-end latency is the more appropriate
deployment quantity.

%% file: Sections/Appendix_C.tex
\section{Extended Results and Analysis}
\label{app:extended_results}

This appendix complements the main-paper results without reproducing their
figures or tables. Main-paper Fig.~3 and Table~1 report the headline
rate--distortion comparison, while main-paper Fig.~4 and Table~2 report the
backbone, antenna-transfer, adaptive-rate, physical-diagnostic, and
Taylor-ablation results. We provide only additional curves, controls, and
analyses not shown there. All experiments use the same disjoint splits as the
main paper, including 1,500 held-out test channels per environment. Training,
model selection, and operating-point selection use label-free quantities; path
annotations are used only for the post-hoc analysis in
Sec.~\ref{app:multipath_fidelity}. We use NMSE throughout. An unquantized GCNO
payload contains $4K$ transmitted real values for $K$ retained paths.

\subsection{Complete Rate--Distortion Curves}
\label{app:complete_rd}

\begin{figure*}[!t]
    \centering
    \includegraphics[width=0.82\textwidth]{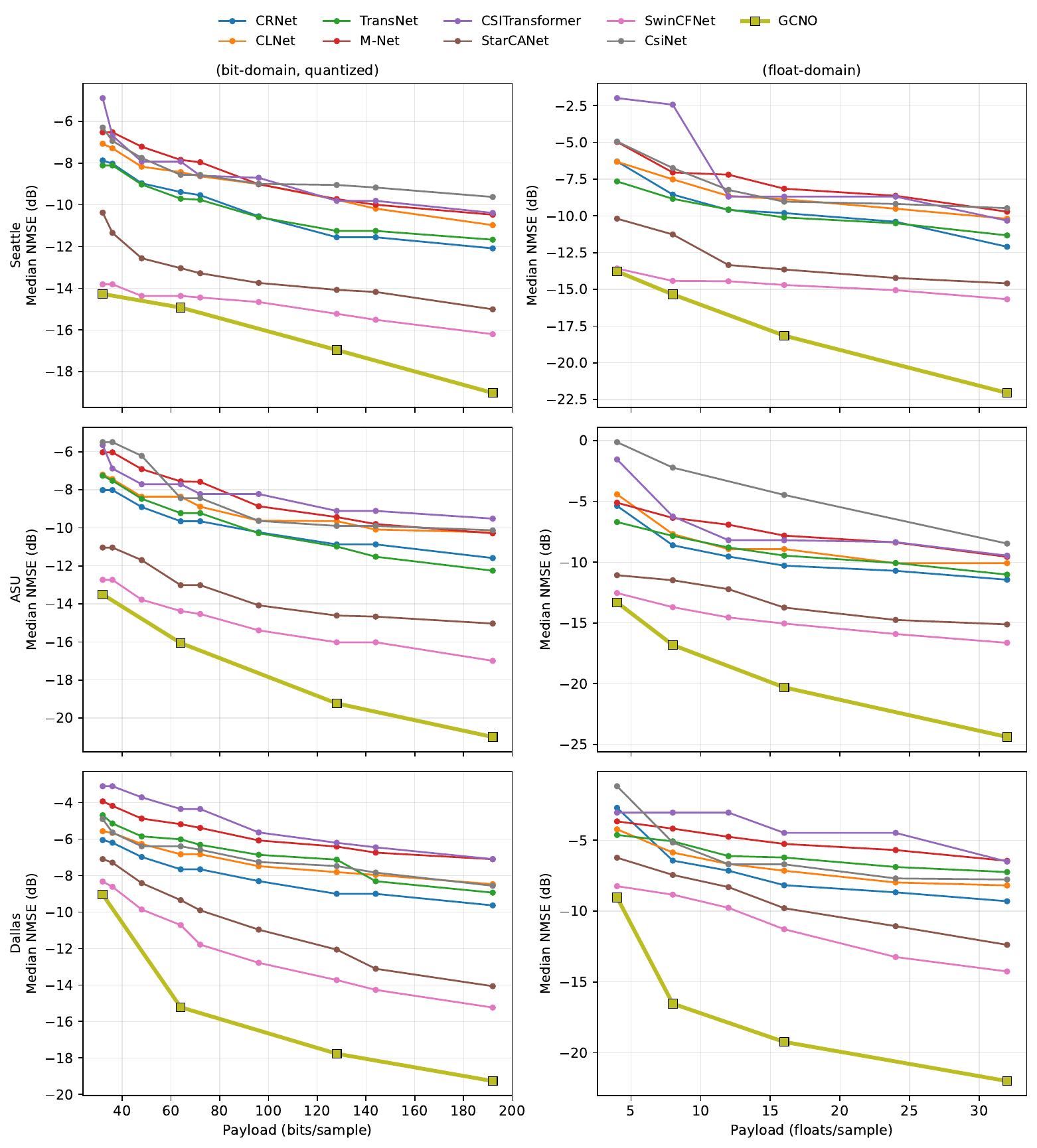}
    \caption{Complete cross-scene rate--distortion curves. Rows correspond to
    Seattle, ASU, and Dallas; columns correspond to quantized bits and transmitted real values. This figure extends main-paper Fig.~3 and Table~1 rather than repeating their selected operating points. Lower and farther left is better.}
    \label{fig:app_all_rd}
\end{figure*}

Main-paper Fig.~3 gives the complete ASU comparison, and main-paper Table~1
gives representative Seattle and Dallas operating points.
Figure~\ref{fig:app_all_rd} adds the complete Seattle and Dallas curves for all
eight paired encoder--decoder baselines. The ASU row is retained only to place
the three environments under one legend and visual scale, rather than as a
separate repetition of the main-paper analysis. The broad ordering reported in
the main paper persists across the full curves: GCNO is strongest over most of
the evaluated range and separates more clearly at medium and larger payloads.
Seattle is the closest comparison at the smallest unquantized budget,
motivating the targeted seed analysis in
Sec.~\ref{app:seattle_seed_stability}. The agreement across three distinct
propagation maps indicates that the observed rate--distortion behavior is not
specific to ASU.

\paragraph{Trainable parameter counts.}
Table~\ref{tab:baseline_parameter_counts} reports the number
of trainable scalar parameters for one representative
instantiated model of each evaluated method. Because some
paired autoencoders change size with their instantiated input
or bottleneck dimensions, these values characterize the
reported model instances rather than configuration-independent
architecture constants. Parameter count is provided only as
model-size context and is distinct from transmitted payload.

\begin{table}[t]
\centering
\caption{Trainable parameter counts for representative
instantiated models.}
\label{tab:baseline_parameter_counts}
\begin{tabular}{lr}
\hline
Method & Trainable parameters \\
\hline
CsiNet         & 302{,}956 \\
CRNet          & 450{,}150 \\
CLNet          & 449{,}270 \\
TransNet       & 1{,}026{,}224 \\
MNet           & 445{,}200 \\
CSITransformer & 563{,}520 \\
StarCANet      & 459{,}688 \\
SwinCFNet      & 10{,}956{,}544 \\
GCNO           & 95{,}253 \\
\hline
\end{tabular}
\end{table}

\subsection{Classical Grid Recovery and Extended Taylor Analysis}
\label{app:classical_taylor}

\paragraph{Grid-OMP control.}
We compare GCNO with Grid-OMP \cite{pati1993omp}, which
selects receive--transmit atoms from the same fixed direction
grid and refits their complex gains by LS. Its stopping
profiles are fixed on validation data, and its payload follows the same $4K$
real-value accounting. Table~\ref{tab:app_omp} shows that Grid-OMP improves
monotonically as more atoms are retained but remains substantially less
accurate under this protocol. Near 16 transmitted values, it obtains
$-4.582$~dB NMSE at 14.949 mean values, whereas the fixed-four-path GCNO
control obtains $-16.511$~dB at 16 values. This GCNO row deliberately disables adaptive-$K$ operation:
every channel is forced to retain exactly $K=4$ paths, giving an
identical 16-value payload for every sample. It is therefore
distinct from the adaptive ASU operating point with a mean
16-value payload in main-paper Table~2, which obtains
$-20.08$~dB. The fixed-$K$ restriction intentionally lowers
GCNO's attainable accuracy to $-16.511$~dB and thereby makes
the comparison more favorable to Grid-OMP; the two reported
GCNO values correspond to different evaluation protocols and
are not contradictory. This diagnostic indicates that the
fixed dictionary and LS fit alone do not explain GCNO's result; it is not
intended as a general ranking of all sparse-recovery algorithms.

\paragraph{Runtime and comparison scope.}
NR configures CSI reporting in slot units and specifies UE
CSI-computation delays in OFDM symbols; the corresponding
wall-clock intervals depend on the numerology and report
configuration \cite{3gpp38331,3gpp38211,3gpp38214}.
These requirements motivate low UE-side processing latency,
but they do not impose a universal inference deadline.
In our batch-1 H100 timing study, Grid-OMP required
approximately 67~ms per channel. For reference, Table~9
reports 15.425~ms for GCNO's mid profile and
20.690--23.830~ms for the approximately 16-value profiles
closest to the comparison in Table~10. We therefore include
Grid-OMP as a lightweight greedy grid-based control rather
than claiming that it is universally the fastest classical
method. More elaborate off-grid procedures can add repeated
detection and refinement steps, as in NOMP
\cite{mamandipoor2016nomp}; a fair broader comparison
would require implementation- and hardware-matched
profiling. Their omission is therefore a scope decision, not a
claim that every classical estimator is categorically
unsuitable for latency-constrained CSI feedback.

\begin{table}[t]
\centering
\small
\caption{Grid-OMP operating points and a fixed-four-path GCNO control.}
\label{tab:app_omp}
\begin{tabular}{l|cc}
\hline
Method & Mean values & NMSE (dB) \\
\hline
Grid-OMP @ 4 values  & 4.000  & $-2.607$ \\
Grid-OMP @ 8 values  & 8.000  & $-3.791$ \\
Grid-OMP @ 16 values & 14.949 & $-4.582$ \\
Grid-OMP @ 32 values & 32.000 & $-5.152$ \\
GCNO ($K=4$)         & 16.000 & $\mathbf{-16.511}$ \\
\hline
\end{tabular}
\end{table}

\paragraph{Taylor refinement beyond the main operating point.}
Main-paper Table~2 already reports the absolute Taylor on/off results at 16
transmitted values, so we do not reproduce them here. Instead,
Table~\ref{tab:app_taylor_sweep} reports the additional NMSE improvement at 4,
8, and 32 values, using a separately retrained no-Taylor control with the same
support-selection and payload rules. The improvement is positive in all nine
additional comparisons and is at least 6.628~dB. These measurements show that
the off-grid correction remains useful across the evaluated payload range; we
do not interpret the variation with payload as a general scaling law.

\begin{table}[t]
\centering
\caption{NMSE improvement from Taylor refinement beyond the 16-value result in
main-paper Table~2. Positive values favor Taylor refinement.}
\label{tab:app_taylor_sweep}
\begin{tabular}{l|ccc}
\hline
Environment & 4 values & 8 values & 32 values \\
\hline
ASU     & 10.729 & 13.649 & 20.439 \\
Dallas  &  6.628 & 13.672 & 17.732 \\
Seattle & 10.819 & 12.176 & 18.529 \\
\hline
\end{tabular}
\end{table}

\subsection{Physical Multipath Fidelity}
\label{app:multipath_fidelity}

\begin{figure}[t]
    \centering
    \includegraphics[width=\columnwidth]{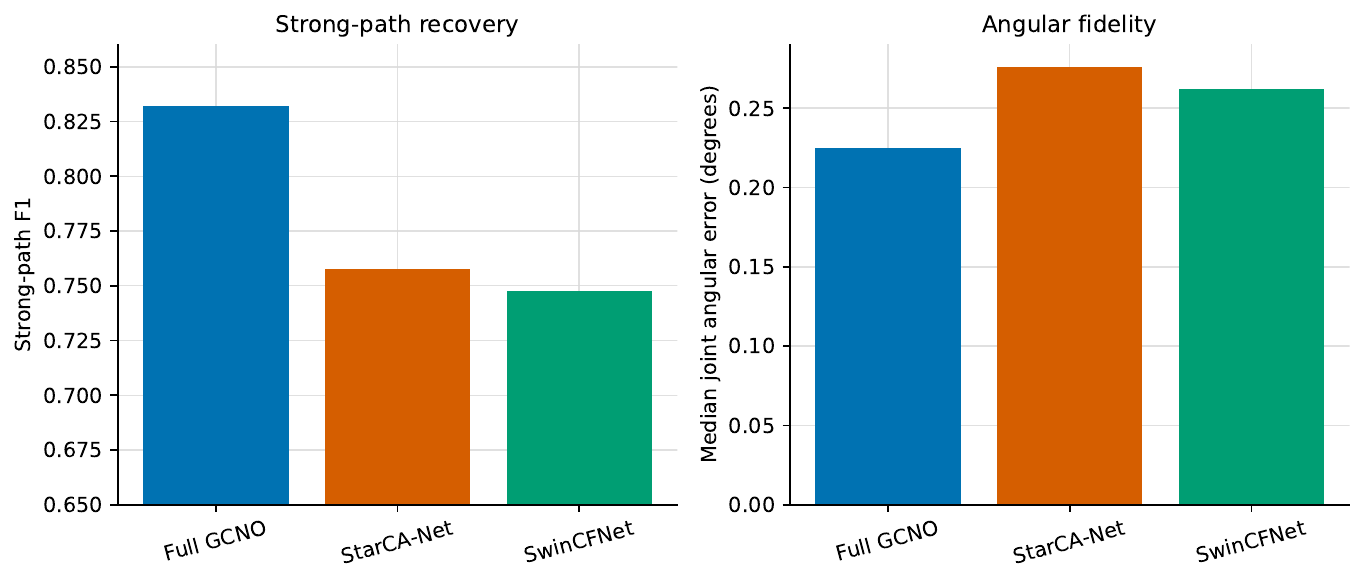}
    \caption{Post-hoc multipath fidelity. Strong-path F1 measures joint support recovery, while median joint effective-angle error measures directional accuracy. Higher is better for F1 and lower is better for angular error.}
    \label{fig:app_multipath_fidelity}
\end{figure}

Main-paper Fig.~4(c) already shows how the selected path count changes with the
requested accuracy, and main-paper Fig.~4(d) reports GCNO's behavior as the
number of strong diagnostic paths increases. We therefore do not repeat those
plots. Figure~\ref{fig:app_multipath_fidelity} instead asks whether the final
reconstructions preserve dominant multipath structure relative to the two
closest neural baselines.

A diagnostic path is considered strong when its amplitude is within 10~dB of
the strongest annotated path in the channel. Receive--transmit direction pairs
are matched jointly within $3^\circ$. GCNO is evaluated directly from its
transmitted tuples. Because StarCANet and SwinCFNet output dense channels rather
than paths, a single frozen offline path probe is applied to their reconstructed
channels solely for this diagnostic; it does not change their transmitted
payloads or NMSE. Under this protocol, GCNO reaches 0.832 strong-path F1 and
$0.225^\circ$ median joint effective-angle error, compared with 0.757 and
$0.276^\circ$ for StarCANet and 0.748 and $0.262^\circ$ for SwinCFNet. GCNO's
weighted strong-path recall is 0.916. These results support the narrower conclusion that
the compact tuple representation preserves dominant resolvable structure well
under this matching protocol; they do not imply recovery of every annotated
ray or that the neural baselines internally represent channels as paths.

\subsection{Protocol for the Array-Size Results in Main-Paper Fig.~4(b)}
\label{app:array_transfer}

Main-paper Fig.~4(b) reports the complete numerical comparison; here we
specify the corresponding evaluation protocol. The experiment uses seven
receive--transmit array configurations:
$16\times16$, $16\times32$, $32\times16$, $24\times24$,
$32\times32$, $32\times64$, and $48\times48$. The renderings are paired so
that a given test index represents the same physical link at every array
shape. Each shape uses the same locked split of 10,000 training, 2,000
validation, and 1,500 test channels. Evaluation is performed on the ordered,
unshuffled 1,500-channel test split. No target-shape test sample is used for
training, fine-tuning, calibration, or operating-point selection.

\paragraph{GCNO transfer.}
The GCNO checkpoint trained at $32\times32$ is applied directly to every target
shape without updating any learned parameter. For each $(N_r,N_t)$, we
recompute the fixed steering and derivative dictionaries, matched-filter
evidence, sample-specific receive and transmit Gramians, and analytical
reconstruction responses at the target dimensions. The learned feature-mixing
weights, Chebyshev coefficients, support head, and Taylor-offset head remain
unchanged. GCNO is therefore evaluated natively at each target resolution,
rather than by resizing the target channel to the training dimensions.

\paragraph{Frozen neural baselines.}
The eight paired neural baselines---CsiNet, CRNet, CLNet, TransNet, MNet,
CSITransformer, StarCANet, and SwinCFNet---use their existing ASU
$75^\circ$-field-of-view checkpoints trained at $32\times32$ with a fixed
latent length of $M=16$ real values. Each model remains
in its original $32\times32$ form: no layer, learned tensor, running statistic,
latent dimension, or resolution-dependent architectural component is changed.
Models are evaluated in inference mode, and each retains its training-split
RMS normalization
\begin{equation}
g_m =
\left[
\frac{1}{|\mathcal T|}
\sum_{x\in\mathcal T}
\frac{\lVert x\rVert_2^2}{2\cdot32\cdot32}
\right]^{-1/2},
\end{equation}
where $\mathcal T$ is the 10,000-channel training split and $m$ indexes the
baseline. The input is multiplied by $g_m$, and the decoder output is divided
by the same fixed value. No normalization statistic is recomputed at a target
array size.

\paragraph{Common deterministic baseline adapter.}
Let $H_{\mathrm{norm}}\in\mathbb C^{N_r\times N_t}$ be the normalized target
channel and define its centered two-channel angular representation as
\begin{equation}
X_N =
\operatorname{stack}_{\Re,\Im}
\left\{
\operatorname{fftshift}
\left[
\mathcal F_{2,\mathrm{ortho}}(H_{\mathrm{norm}})
\right]
\right\},
N=(N_r,N_t).
\end{equation}
For source shape $a=(a_r,a_t)$ and destination shape $b=(b_r,b_t)$, the
external adapter is
\begin{equation}
\mathcal A_{a\rightarrow b}(X)
=
\sqrt{\frac{a_ra_t}{b_rb_t}}\,
\mathcal B_{a\rightarrow b}(X),
\label{eq:array_adapter}
\end{equation}
where $\mathcal B$ bilinearly interpolates the real and imaginary channels
with \texttt{align\_corners=False}. Antialiasing is enabled whenever a
dimension is downsampled. If $a=b$, the implementation returns an exact clone,
with no interpolation or antialiasing. The square-root factor compensates for
the change in angular-grid size, and the reverse adapter uses its reciprocal.

For baseline $m$, the complete frozen inference path is
\begin{align}
z_m &=
\mathcal E_m\!\left(
g_m\,
\operatorname{ifftshift}
\left[
\mathcal A_{N\rightarrow(32,32)}(X_N)
\right]
\right),\\
\widehat X_N &=
\mathcal A_{(32,32)\rightarrow N}
\left(
\operatorname{fftshift}
\left[
g_m^{-1}\mathcal D_m(z_m)
\right]
\right),
\end{align}
where $\mathcal E_m$ and $\mathcal D_m$ are the unchanged encoder and decoder.
The reconstructed angular representation is finally mapped back to the channel
domain using the inverse centered orthonormal two-dimensional DFT.

Table~\ref{tab:baseline_array_adapter} lists the resulting size factors and
antialiasing operations. ``Forward'' denotes target-to-$32\times32$ adaptation,
and ``reverse'' denotes $32\times32$-to-target adaptation.

\begin{table*}[t]
\centering
\caption{Deterministic input/output adapter used for every frozen neural
baseline in main-paper Fig.~4(b). AA denotes antialiasing. The latent payload
remains 16 real values for every array shape.}
\label{tab:baseline_array_adapter}
\begin{tabular}{c|cc|cc|cc}
\hline
Target shape
& Forward scale & Forward AA
& Reverse scale & Reverse AA
& Payload & Compression \\
\hline
$16\times16$ & $1/2$          & No
             & $2$            & Both dimensions
             & 16 & $32\times$ \\
$16\times32$ & $1/\sqrt{2}$   & No
             & $\sqrt{2}$     & Receive dimension
             & 16 & $64\times$ \\
$32\times16$ & $1/\sqrt{2}$   & No
             & $\sqrt{2}$     & Transmit dimension
             & 16 & $64\times$ \\
$24\times24$ & $3/4$          & No
             & $4/3$          & Both dimensions
             & 16 & $72\times$ \\
$32\times32$ & $1$            & Exact identity
             & $1$            & Exact identity
             & 16 & $128\times$ \\
$32\times64$ & $\sqrt{2}$     & Transmit dimension
             & $1/\sqrt{2}$   & No
             & 16 & $256\times$ \\
$48\times48$ & $3/2$          & Both dimensions
             & $2/3$          & No
             & 16 & $288\times$ \\
\hline
\end{tabular}
\end{table*}

The adapter is deterministic, parameter-free, and carries no side information.
The baseline payload is always 16 real values, giving compression
$2N_rN_t/16$. At the native $32\times32$ shape, both adapter directions are
exact identities, providing the source-resolution control. Thus, the
comparison in main-paper Fig.~4(b) evaluates reuse of frozen models under one
common nonlearned dimension-matching rule, rather than models retrained
separately for each array configuration. Its scope is limited to the seven
tested uniform-linear-array shapes and does not imply invariance to arbitrary
array layouts or calibration changes.

\subsection{Cross-Scene Transfer Without Retraining}
\label{app:cross_scene_transfer}

We further test whether a compressor trained in one propagation
scene remains useful in another (Table \ref{tab:app_cross_scene_transfer}). For each ordered source--target
pair among Dallas, Seattle, and ASU, each model is trained only
on the source scene and evaluated on the held-out target-scene
test split with all learned parameters frozen. No target-scene
retraining or fine-tuning is performed. All methods use the
nominal 16-value operating point. GCNO reconstructs only from
its transmitted path tuples and LS gains, whereas SwinCFNet and
StarCANet use their unchanged paired decoders.

\begin{table}[t]
\centering
\small
\caption{Cross-scene median NMSE (dB) at the nominal
16-value operating point. Lower is better.}
\label{tab:app_cross_scene_transfer}
\begin{tabular}{l|rrr}
\hline
Source $\rightarrow$ target & GCNO & SwinCFNet & StarCANet \\
\hline
Dallas $\rightarrow$ Seattle
    & $\mathbf{-14.203}$ & $-14.037$ & $-11.799$ \\
Dallas $\rightarrow$ ASU
    & $\mathbf{-14.815}$ & $-10.490$ & $-9.747$ \\
Seattle $\rightarrow$ Dallas
    & $\mathbf{-14.197}$ & $-7.146$ & $-6.572$ \\
Seattle $\rightarrow$ ASU
    & $\mathbf{-15.191}$ & $-9.560$ & $-9.522$ \\
ASU $\rightarrow$ Dallas
    & $\mathbf{-15.401}$ & $-6.962$ & $-6.308$ \\
ASU $\rightarrow$ Seattle
    & $\mathbf{-15.821}$ & $-12.932$ & $-11.092$ \\
\hline
\end{tabular}
\end{table}

GCNO obtains the lowest median NMSE in all six transfer
directions, improving over SwinCFNet by $0.166$--$8.439$~dB.
This supports the narrower conclusion that its physics-aligned
support representation and analytical reconstruction transfer
more robustly under the tested scene shifts than the paired
latent-code baselines. The result is consistent with, but does
not by itself prove, reduced dependence on scene-specific
statistics; it does not imply invariance to arbitrary propagation
environments.  

\subsection{Beamforming Utility}
\label{app:beamforming}

We next test whether the reconstructed channel preserves the dominant transmit
subspace used in a standard beamforming calculation. Channels are normalized to
unit Frobenius norm. For one stream, the precoder is the dominant right singular
vector of the reconstructed channel. For two streams, the two dominant right
singular vectors are assigned equal power. In both cases, the receiver uses the
optimal combiner for the original effective channel. Perfect CSI is included
only as a dense reference. GCNO uses 14.636/12/32 mean/median/95th-percentile
transmitted values, whereas each neural comparator uses 16 values for every
channel.

Table~\ref{tab:app_beamforming} reports mean spectral efficiency. At 20~dB,
GCNO obtains 6.5319~bit/s/Hz for one stream, within approximately
0.004~bit/s/Hz of perfect CSI and above both compressed baselines. For two
streams, GCNO obtains 6.8240~bit/s/Hz, compared with 6.9644~bit/s/Hz for
perfect CSI. The squared alignment of the dominant right singular vector has
mean 0.9973, median 0.9995, and fifth percentile 0.9901. These results support
preservation of the dominant transmit subspace under this single-user,
narrowband, optimal-combiner protocol; they do not establish the same behavior
for multiuser scheduling, other precoders, receiver mismatch, or imperfect
channel estimation.

\begin{table}[t]
\centering
\small
\caption{Mean spectral efficiency (bit/s/Hz). Bold marks the strongest
compressed method; perfect CSI is a dense reference.}
\label{tab:app_beamforming}
\begin{tabular}{l|rrrr}
\hline
Method & 0 dB & 10 dB & 20 dB & 30 dB \\
\hline
\multicolumn{5}{c}{One stream} \\
GCNO        & $\mathbf{0.9411}$ & $\mathbf{3.3452}$ & $\mathbf{6.5319}$ & $\mathbf{9.8395}$ \\
StarCANet   & 0.9204 & 3.2864 & 6.4447 & 9.7381 \\
SwinCFNet   & 0.9212 & 3.2875 & 6.4488 & 9.7480 \\
Perfect CSI & 0.9428 & 3.3488 & 6.5360 & 9.8437 \\
\hline
\multicolumn{5}{c}{Two streams} \\
GCNO        & $\mathbf{0.5871}$ & $\mathbf{2.7846}$ & $\mathbf{6.8240}$ & $\mathbf{12.0669}$ \\
StarCANet   & 0.5665 & 2.6629 & 6.3938 & 11.1730 \\
SwinCFNet   & 0.5725 & 2.7050 & 6.5858 & 11.6554 \\
Perfect CSI & 0.5906 & 2.8101 & 6.9644 & 12.4959 \\
\hline
\end{tabular}
\end{table}

\subsection{Seattle Seed Stability under a Baseline-Favorable Envelope}
\label{app:seattle_seed_stability}

Seattle has the smallest separation between GCNO and the strongest neural
baseline in the main comparison. We therefore use it for a targeted
seed-sensitivity check rather than repeating the experiment in every
environment. Three GCNO runs, using seeds 43, 44, and 45, are evaluated on the
same fixed 1,500-channel test split (Table \ref{tab:gcno_seed_stability}.

We construct a deliberately baseline-favorable comparison (Table \ref{tab:baseline_favorable_best_seed}). For each baseline
and nominal payload target, we retain the single lowest NMSE obtained across
its three evaluated seeds. The selected baseline seed may therefore change
independently at each operating point. By contrast, each GCNO seed represents
one trained model evaluated across all three payload targets. Because the
realized adaptive GCNO payloads differ slightly within a nominal target, we
also retain the most favorable baseline value among the corresponding
matched-payload comparisons. This protocol is stricter than either averaging
the baseline seeds or retaining one baseline seed over the full curve.

\begin{table}[t]
\centering
\caption{GCNO seed-level median NMSE across three seeds. All values are in dB.}
\label{tab:gcno_seed_stability}
\scriptsize
\begin{tabular}{
@{}c
@{\hspace{5pt}}r
@{\hspace{5pt}}r
@{\hspace{5pt}}r
@{\hspace{5pt}}c@{}
}
\hline
Payload & Seed 43 & Seed 44 & Seed 45 & $\mu \pm \sigma$ \\
\hline
8
& $-15.5861$
& $-15.0772$
& $-15.1942$
& $-15.2858 \pm 0.2666$ \\
16
& $-18.6382$
& $-17.1041$
& $-17.6077$
& $-17.7833 \pm 0.7820$ \\
32
& $-22.4733$
& $-20.3709$
& $-19.9455$
& $-20.9299 \pm 1.3534$ \\
\hline
\end{tabular}
\end{table}

\begin{table}[t]
\centering
\caption{Baseline-favorable best-seed results at each target payload. All values are in dB.}
\label{tab:baseline_favorable_best_seed}
\scriptsize
\begin{tabular}{
@{}c
@{\hspace{7pt}}l
@{\hspace{7pt}}r
@{\hspace{7pt}}r@{}
}
\hline
Target & Comparator & Best NMSE & $\min \Delta$ \\
\hline
8  & SwinCFNet & $-14.4309$ & $+0.6463$ \\
8  & StarCANet & $-12.0964$ & $+2.9808$ \\
\hline
16 & SwinCFNet & $-14.7013$ & $+2.4028$ \\
16 & StarCANet & $-14.0868$ & $+3.0173$ \\
\hline
32 & SwinCFNet & $-15.7820$ & $+4.1635$ \\
32 & StarCANet & $-14.7140$ & $+5.2315$ \\
\hline
\end{tabular}
\end{table}

Here,
$\Delta=\mathrm{NMSE}_{\mathrm{baseline}}-\mathrm{NMSE}_{\mathrm{GCNO}}$,
so a positive value indicates lower NMSE for GCNO, and $\min\Delta$ is the
smallest margin over the three GCNO runs. The ordering remains unchanged for
every GCNO seed, payload target, and comparator, despite allowing each baseline
to use its strongest seed independently at every operating point. The smallest
observed margin is $0.6463$~dB over SwinCFNet and $2.9808$~dB over StarCANet.
At the 16- and 32-value targets, the minimum margins over SwinCFNet increase to
$2.4028$ and $4.1635$~dB, respectively. This targeted check indicates that the
Seattle ordering is not attributable to one favorable GCNO initialization
within the evaluated runs; it is not intended as an exhaustive characterization
of all possible random seeds.

%% file: aaai2027.bib
@techreport{3gpp38331,
  author      = {{3GPP}},
  title       = {{NR; Radio Resource Control (RRC) Protocol Specification}},
  institution = {European Telecommunications Standards Institute},
  number      = {ETSI TS 138 331 V18.6.0
                 (3GPP TS 38.331, Release 18)},
  year        = {2025},
  note        = {Sec. 6.3.2,
                 CSI-ReportPeriodicityAndOffset}
}

@techreport{3gpp38211,
  author      = {{3GPP}},
  title       = {{NR; Physical Channels and Modulation}},
  institution = {European Telecommunications Standards Institute},
  number      = {ETSI TS 138 211 V18.7.0
                 (3GPP TS 38.211, Release 18)},
  year        = {2025},
  note        = {Sec. 4.3.2 and Table 4.3.2-1}
}

@techreport{3gpp38214,
  author      = {{3GPP}},
  title       = {{NR; Physical Layer Procedures for Data}},
  institution = {European Telecommunications Standards Institute},
  number      = {ETSI TS 138 214 V18.7.0
                 (3GPP TS 38.214, Release 18)},
  year        = {2025},
  note        = {Sec. 5.4 and Tables 5.4-1--5.4-2}
}

@inproceedings{pati1993omp,
  author    = {Pati, Y. C. and Rezaiifar, R. and
               Krishnaprasad, P. S.},
  title     = {Orthogonal Matching Pursuit: Recursive Function
               Approximation with Applications to Wavelet
               Decomposition},
  booktitle = {Proceedings of the 27th Asilomar Conference on
               Signals, Systems and Computers},
  pages     = {40--44},
  year      = {1993},
  doi       = {10.1109/ACSSC.1993.342465}
}

@article{mamandipoor2016nomp,
  author  = {Mamandipoor, Babak and Ramasamy, Dinesh and
             Madhow, Upamanyu},
  title   = {Newtonized Orthogonal Matching Pursuit:
             Frequency Estimation over the Continuum},
  journal = {IEEE Transactions on Signal Processing},
  volume  = {64},
  number  = {19},
  pages   = {5066--5081},
  year    = {2016},
  doi     = {10.1109/TSP.2016.2580523}
}

@article{alkhateeb2019deepmimo,
  title={DeepMIMO: A generic deep learning dataset for millimeter wave and massive MIMO applications},
  author={Alkhateeb, Ahmed},
  journal={arXiv preprint arXiv:1902.06435},
  year={2019}
}

@article{klukas1998line,
  title={Line-of-sight angle of arrival estimation in the outdoor multipath environment},
  author={Klukas, Richard and Fattouche, Michel},
  journal={IEEE transactions on vehicular technology},
  volume={47},
  number={1},
  pages={342--351},
  year={1998},
  publisher={IEEE}
}

@article{ji2021clnet,
  author  = {Ji, Sijie and Li, Mo},
  title   = {{CLNet}: Complex Input Lightweight Neural Network Designed for Massive {MIMO} {CSI} Feedback},
  journal = {IEEE Wireless Communications Letters},
  year    = {2021},
  volume  = {10},
  number  = {10},
  pages   = {2318--2322},
  month   = oct,
  doi     = {10.1109/LWC.2021.3100493}
}

@inproceedings{yu2023mnet,
  author    = {Yu, Yaxin and Teng, Yinglei and Wang, Binghui and Liu, An and Lau, Vincent},
  title     = {{M-Net}: A Lightweight Network Based on Multilayer Perceptron for Massive {MIMO} {CSI} Feedback},
  booktitle = {2023 IEEE Globecom Workshops (GC Wkshps)},
  year      = {2023},
  pages     = {26--31},
  publisher = {IEEE},
  doi       = {10.1109/GCWKSHPS58843.2023.10464906}
}

@article{zhao2026starcanet,
  author  = {Zhao, Kai and Wu, Haiyi and Xiong, Yong and Zhu, Leiji and Xu, Minhao},
  title   = {{StarCANet}: A Compact and Efficient Neural Network for Massive {MIMO} {CSI} Feedback},
  journal = {IEEE Wireless Communications Letters},
  year    = {2026},
  volume  = {15},
  pages   = {540--544},
  doi     = {10.1109/LWC.2025.3631339}
}

@inproceedings{cheng2023swincfnet,
  author    = {Cheng, Jiaming and Chen, Wei and Xu, Jialong and Guo, Yiran and Li, Lun and Ai, Bo},
  title     = {Swin Transformer-Based {CSI} Feedback for Massive {MIMO}},
  booktitle = {2023 IEEE 23rd International Conference on Communication Technology (ICCT)},
  year      = {2023},
  pages     = {809--814},
  publisher = {IEEE},
  doi       = {10.1109/ICCT59356.2023.10419637}
}

@article{elayach2014spatially,
  author  = {El Ayach, Omar and Rajagopal, Sridhar and Abu-Surra, Shadi and Pi, Zhouyue and Heath, Robert W.},
  title   = {Spatially Sparse Precoding in Millimeter Wave {MIMO} Systems},
  journal = {IEEE Transactions on Wireless Communications},
  volume  = {13},
  number  = {3},
  pages   = {1499--1513},
  year    = {2014},
  doi     = {10.1109/TWC.2014.011714.130846}
}

@article{akdeniz2014millimeter,
  author  = {Akdeniz, Mustafa Riza and Liu, Yuanpeng and Samimi, Mathew K. and Sun, Shu and Rangan, Sundeep and Rappaport, Theodore S. and Erkip, Elza},
  title   = {Millimeter Wave Channel Modeling and Cellular Capacity Evaluation},
  journal = {IEEE Journal on Selected Areas in Communications},
  volume  = {32},
  number  = {6},
  pages   = {1164--1179},
  year    = {2014},
  doi     = {10.1109/JSAC.2014.2328154}
}

@article{csinet,
  author = {Wen, Chao-Kai and Shih, Wan-Ting and Jin, Shi},
  title = {{Deep Learning for Massive MIMO CSI Feedback}},
  journal = {IEEE Wireless Communications Letters},
  volume = {7},
  number = {5},
  pages = {748--751},
  year = {2018},
  doi = {10.1109/LWC.2018.2818160}
}

@inproceedings{crnet,
  author = {Lu, Zhilin and Wang, Jintao and Song, Jian},
  title = {{Multi-Resolution CSI Feedback With Deep Learning in Massive MIMO System}},
  booktitle = {ICC 2020 -- 2020 IEEE International Conference on Communications (ICC)},
  pages = {1--6},
  year = {2020},
  doi = {10.1109/ICC40277.2020.9149229}
}

@article{csinet_plus,
  author = {Guo, Jiajia and Wen, Chao-Kai and Jin, Shi and Li, Geoffrey Ye},
  title = {{Convolutional Neural Network-Based Multiple-Rate Compressive Sensing for Massive MIMO CSI Feedback: Design, Simulation, and Analysis}},
  journal = {IEEE Transactions on Wireless Communications},
  volume = {19},
  number = {4},
  pages = {2827--2840},
  year = {2020},
  doi = {10.1109/TWC.2020.2968430}
}

@article{dcrnet,
  author = {Tang, Shunpu and Xia, Junjuan and Fan, Lisheng and Lei, Xianfu and Xu, Wei and Nallanathan, Arumugam},
  title = {{Dilated Convolution Based CSI Feedback Compression for Massive MIMO Systems}},
  journal = {IEEE Transactions on Vehicular Technology},
  volume = {71},
  number = {10},
  pages = {11216--11221},
  year = {2022},
  doi = {10.1109/TVT.2022.3183596}
}

@article{csinet_lstm,
  author = {Wang, Tianqi and Wen, Chao-Kai and Jin, Shi and Li, Geoffrey Ye},
  title = {{Deep Learning-Based CSI Feedback Approach for Time-Varying Massive MIMO Channels}},
  journal = {IEEE Wireless Communications Letters},
  volume = {8},
  number = {2},
  pages = {416--419},
  year = {2019}
}

@article{attention_csi,
  author = {Li, Qi and Zhang, Aihua and Liu, Pengcheng and Li, Jianjun and Li, Chunlei},
  title = {{A Novel CSI Feedback Approach for Massive MIMO Using LSTM-Attention CNN}},
  journal = {IEEE Access},
  volume = {8},
  pages = {7295--7302},
  year = {2020},
  doi = {10.1109/ACCESS.2020.2963896}
}

@article{transnet,
  author = {Cui, Yaodong and Guo, Aihuang and Song, Chunlin},
  title = {{TransNet: Full Attention Network for CSI Feedback in FDD Massive MIMO System}},
  journal = {IEEE Wireless Communications Letters},
  volume = {11},
  number = {5},
  pages = {903--907},
  year = {2022},
  doi = {10.1109/LWC.2022.3149416}
}

@inproceedings{fno,
  author = {Li, Zongyi and Kovachki, Nikola and Azizzadenesheli, Kamyar and Liu, Burigede and Bhattacharya, Kaushik and Stuart, Andrew and Anandkumar, Anima},
  title = {{Fourier Neural Operator for Parametric Partial Differential Equations}},
  booktitle = {International Conference on Learning Representations},
  year = {2021},
  url = {https://openreview.net/forum?id=c8P9NQVtmnO}
}

@article{deeponet,
  author = {Lu, Lu and Jin, Pengzhan and Pang, Guofei and Zhang, Zhongqiang and Karniadakis, George Em},
  title = {{Learning Nonlinear Operators via DeepONet Based on the Universal Approximation Theorem of Operators}},
  journal = {Nature Machine Intelligence},
  volume = {3},
  number = {3},
  pages = {218--229},
  year = {2021},
  doi = {10.1038/s42256-021-00302-5}
}

@inproceedings{chebnet,
  author = {Defferrard, Micha{\"e}l and Bresson, Xavier and Vandergheynst, Pierre},
  title = {{Convolutional Neural Networks on Graphs with Fast Localized Spectral Filtering}},
  booktitle = {Advances in Neural Information Processing Systems 29},
  pages = {3837--3845},
  year = {2016}
}

@inproceedings{omp,
  author = {Pati, Yagyensh Chandra and Rezaiifar, Ramin and Krishnaprasad, P. S.},
  title = {{Orthogonal Matching Pursuit: Recursive Function Approximation with Applications to Wavelet Decomposition}},
  booktitle = {Conference Record of the Twenty-Seventh Asilomar Conference on Signals, Systems and Computers},
  volume = {1},
  pages = {40--44},
  year = {1993},
  doi = {10.1109/ACSSC.1993.342465}
}

@article{basis_pursuit,
  author = {Chen, Scott Shaobing and Donoho, David L. and Saunders, Michael A.},
  title = {{Atomic Decomposition by Basis Pursuit}},
  journal = {SIAM Journal on Scientific Computing},
  volume = {20},
  number = {1},
  pages = {33--61},
  year = {1998},
  doi = {10.1137/S1064827596304010}
}

@article{sbl,
  author = {Tipping, Michael E.},
  title = {{Sparse Bayesian Learning and the Relevance Vector Machine}},
  journal = {Journal of Machine Learning Research},
  volume = {1},
  pages = {211--244},
  year = {2001}
}

@article{esprit,
  author = {Roy, Richard and Kailath, Thomas},
  title = {{{ESPRIT}---Estimation of Signal Parameters via Rotational Invariance Techniques}},
  journal = {IEEE Transactions on Acoustics, Speech, and Signal Processing},
  volume = {37},
  number = {7},
  pages = {984--995},
  year = {1989},
  doi = {10.1109/29.32276}
}

@article{offgrid_sbl,
  author = {Yang, Zai and Xie, Lihua and Zhang, Cishen},
  title = {{Off-Grid Direction of Arrival Estimation Using Sparse Bayesian Inference}},
  journal = {IEEE Transactions on Signal Processing},
  volume = {61},
  number = {1},
  pages = {38--43},
  year = {2013},
  doi = {10.1109/TSP.2012.2222378}
}

@article{newton_refine,
  author = {Mamandipoor, Babak and Ramasamy, Dinesh and Madhow, Upamanyu},
  title = {{Newtonized Orthogonal Matching Pursuit: Frequency Estimation Over the Continuum}},
  journal = {IEEE Transactions on Signal Processing},
  volume = {64},
  number = {19},
  pages = {5066--5081},
  year = {2016},
  doi = {10.1109/TSP.2016.2582161}
}

@inproceedings{lista,
  author = {Gregor, Karol and LeCun, Yann},
  title = {{Learning Fast Approximations of Sparse Coding}},
  booktitle = {Proceedings of the 27th International Conference on Machine Learning},
  pages = {399--406},
  year = {2010}
}

@article{unrolled_sparse,
  author = {Borgerding, Mark and Schniter, Philip and Rangan, Sundeep},
  title = {{{AMP}-Inspired Deep Networks for Sparse Linear Inverse Problems}},
  journal = {IEEE Transactions on Signal Processing},
  volume = {65},
  number = {16},
  pages = {4293--4308},
  year = {2017},
  doi = {10.1109/TSP.2017.2708040}
}

@article{marzetta2010,
  author  = {Marzetta, Thomas L.},
  title   = {Noncooperative Cellular Wireless with Unlimited Numbers of Base Station Antennas},
  journal = {IEEE Transactions on Wireless Communications},
  volume  = {9},
  number  = {11},
  pages   = {3590--3600},
  year    = {2010}
}

@article{larsson2014,
  author  = {Larsson, Erik G. and Edfors, Ove and Tufvesson, Fredrik and Marzetta, Thomas L.},
  title   = {Massive {MIMO} for Next Generation Wireless Systems},
  journal = {IEEE Communications Magazine},
  volume  = {52},
  number  = {2},
  pages   = {186--195},
  year    = {2014}
}

@article{oshea2017,
  author  = {O'Shea, Timothy and Hoydis, Jakob},
  title   = {An Introduction to Deep Learning for the Physical Layer},
  journal = {IEEE Transactions on Cognitive Communications and Networking},
  volume  = {3},
  number  = {4},
  pages   = {563--575},
  year    = {2017}
}

@article{guo2022overview,
  title={Overview of deep learning-based CSI feedback in massive MIMO systems},
  author={Guo, Jiajia and Wen, Chao-Kai and Jin, Shi and Li, Geoffrey Ye},
  journal={IEEE Transactions on Communications},
  volume={70},
  number={12},
  pages={8017--8045},
  year={2022},
  publisher={IEEE}
}

@inproceedings{xu2021transformer,
  author    = {Xu, Yikun and Yuan, Ming and Pun, Man-On},
  title     = {Transformer Empowered {CSI} Feedback for Massive {MIMO} Systems},
  booktitle = {2021 30th Wireless and Optical Communications Conference (WOCC)},
  pages     = {157--161},
  year      = {2021}
}

@article{alkhateeb2014,
  author  = {Alkhateeb, Ahmed and El Ayach, Omar and Leus, Geert and Heath, Robert W.},
  title   = {Channel Estimation and Hybrid Precoding for Millimeter Wave Cellular Systems},
  journal = {IEEE Journal of Selected Topics in Signal Processing},
  volume  = {8},
  number  = {5},
  pages   = {831--846},
  year    = {2014}
}

@inproceedings{wagle2025physics,
  title={Physics-based generative models for geometrically consistent and interpretable wireless channel synthesis},
  author={Wagle, Satyavrat and Malhotra, Akshay and Hamidi-Rad, Shahab and Sant, Aditya and Love, David J and Brinton, Christopher G},
  booktitle={Proceedings of the Thirty-Fourth International Joint Conference on Artificial Intelligence},
  pages={9384--9392},
  year={2025}
}

@inproceedings{wagle2025physics_workshop,
  title={Physics-Informed Generative Approaches for Wireless Channel Modeling},
  author={Wagle, Satyavrat and Malhotra, Akshay and Hamidi-Rad, Shahab and Sant, Aditya and Love, David J and Brinton, Christopher G},
  booktitle={ICLR 2025 Workshop on Deep Generative Model in Machine Learning: Theory, Principle and Efficacy},
  year={2025}
}

@article{tang2013offgrid,
  author  = {Tang, Gongguo and Bhaskar, Badri Narayan and Shah, Parikshit and Recht, Benjamin},
  title   = {Compressed Sensing Off the Grid},
  journal = {IEEE Transactions on Information Theory},
  volume  = {59},
  number  = {11},
  pages   = {7465--7490},
  year    = {2013}
}
